\documentclass[10pt, conference, letterpaper]{IEEEtran}

\def\ps@headings{%
\def\@oddhead{\mbox{}\scriptsize\rightmark \hfil \thepage}%
\def\@evenhead{\scriptsize\thepage \hfil \leftmark\mbox{}}%
\def\@oddfoot{}%
\def\@evenfoot{}}
\makeatother 

\usepackage{cite}
\usepackage{amsfonts}
\usepackage{graphics}
\usepackage{graphicx,color}
\usepackage{textcomp}
\usepackage{epsfig}
\usepackage{latexsym}
\usepackage{amsmath}
\usepackage{amssymb}
\usepackage{subfigure}
\usepackage{array}
\usepackage{url}
\usepackage{algorithm}
\usepackage{algorithmic}
\usepackage{subfigure}
\usepackage{amsmath}
\usepackage{balance}

\DeclareMathOperator*{\argmin}{arg\,min}

\begin{document}

\title{On the Detection of Adaptive Adversarial Attacks in Speaker Verification Systems}

\author{Zesheng Chen\\
Department of Computer Science \\
Purdue University Fort Wayne, Indiana 46805 \\
Email: chenz@pfw.edu }

\maketitle

\begin{abstract}
Speaker verification systems have been widely used in smart phones and Internet of things devices
to identify legitimate users.
In recent work, it has been shown that adversarial attacks, such as FAKEBOB,
can work effectively against speaker verification systems.
The goal of this paper is to design a detector
that can distinguish an original audio from an audio contaminated by adversarial attacks.
Specifically, our designed detector, called {\em MEH-FEST}, calculates
the minimum energy in high frequencies from the short-time Fourier transform of an audio
and uses it as a detection metric.
Through both analysis and experiments, we show that our proposed detector is easy to implement,
fast to process an input audio, and effective in determining whether an audio is corrupted by FAKEBOB attacks.
The experimental results indicate that the detector is extremely effective:
with near zero false positive and false negative rates for detecting FAKEBOB attacks in
Gaussian mixture model (GMM) and i-vector speaker verification systems.
Moreover, adaptive adversarial attacks against our proposed detector and their countermeasures are discussed and studied,
showing the game between attackers and defenders.
\end{abstract}


\section{Introduction}

With the popularity of smart phones and Internet of things (IoT) devices at home ({\it e.g.},
Amazon Alexa, Apple Siri, and Google Assistant), voice control becomes a main interface between humans and devices,
because of its convenience and ease of operation.
To secure such an interface, speaker verification systems have been widely applied to verify a user's identity
through their voice before allowing them to access a device.
Moreover, some e-banking systems have also authorized visitors through their voices.
In other words, human voice has been used as biometrics to distinguish between legitimate users and illegal users.

Two main security attacks have recently emerged to tamper with speaker verification systems.
One is called {\em replay attacks} that record the legitimate user's speech and then replay it to fool a speaker verification system \cite{Wang}.
Such a sniffing and spoofing attack requires an attacker to obtain a legitimate user's audio.
The other attack is called {\em adversarial attacks} that generate a speech acceptable to the system
by adding small and well-designed perturbations to an illegal user's speech \cite{Chen,Chang}.
Such an attack does not need a copy of the legitimate user's speech and is imperceptible to humans.
In this work, we focus on adversarial attacks.

Adversarial attacks were discovered when machine learning classifiers were applied at test time in adversarial setting \cite{Dalvi,Lowd,Biggio}.
With the popularity of deep learning, Szegedy et al. showed that image classifiers are particularly vulnerable to adversarial attacks \cite{Szegedy, Goodfellow}.
Moreover, it has been found that adversarial attacks can be applied to a wide range of domains
such as image steganography \cite{Tang}, multimedia forensics \cite{Chen_C}, and malware detection \cite{Li_D}.
Since the currently widely-used speaker verification systems, such as GMM, i-vector, d-vector, and x-vector,
are based on machine learning, they are vulnerable to adversarial attacks.
Specially, Chen et al. designed a black-box adversarial attack, called {\em FAKEBOB},
that does not require the implementation information of a speaker verification system
and only needs the output scores of the system \cite{Chen}.
FAKEBOB is shown to be very effective against both open-source and commercial systems,
and can achieve 99\% targeted attack success rate.
Moreover, it has been shown in \cite{Chen} that
several defense methods, including local smoothing, quantization, and temporal dependency detection \cite{Yang},
that work well against adversarial attacks in the image domain are not able to counteract FAKEBOB.

There have been some recent works proposed to defend against adversarial attacks
in speaker verification systems \cite{Li,Wu,Chang,Villalba,Joshi,Abdullah,Jati,Das}.
For example, Li et al. used a separate neural network to detect adversarial samples \cite{Li}.
Joshi et al. studied generative adversarial networks (GAN) based and variational autoencoders (VAE) based defenses \cite{Joshi}.
Wu et al. proposed to sample the neighbors of a given speech and calculate the average score based on these neighbors \cite{Wu}.
In our previous work, we designed a defense system by adding small random Gaussian noise to an input audio \cite{Chang}.
However, the following research question still remains:
How can we effectively and efficiently distinguish between an original audio and an adversarial audio?
The original audio is either from a legitimate user or from an illegal user and is without perturbations,
whereas the adversarial audio can be either a successful attack or a failed attack
and is with the attacker's designed perturbations.

Specifically, we consider the following scenario:
If an illegal audio, before contaminated by an adversarial attack, has a similar signal-to-noise ratio (SNR) as the legitimate audio,
how can we accurately detect the adversarial audio in real time?
The goal of this work is to design a detector that is simple, fast, and effective in determining
whether a given audio is from an adversarial attack such as FAKEBOB.
To achieve this goal, we propose a new detector called {\it Minimum Energy in High FrEquencies for Short Time (MEH-FEST)}.
We give such a name in hope to bring disappointment ({\it i.e.}, MEH) to attackers and to focus on the most important
area of an audio signal ({\it i.e.}, FEST).
Our designed MEH-FEST detector is based on the following three key observations:
\begin{itemize}
\item
Although small, perturbations in adversarial attacks against speaker verification systems behave like white noise and
appear everywhere in the audio signal across time and frequency.

\item
Audio signals are non-stationary.
As a result, the way that perturbations affect the audio signal is significantly different
when a speech is present or absent.

\item
The energy in high frequencies in an original audio is usually small, especially when a speech is absent.
\end{itemize}

Specifically, the MEH-FEST detector applies short-time Fourier transform \cite{Muller} and
calculates the minimum energy of an audio signal in high frequencies among short-time periods.
Through analysis and experiments, we demonstrate that our proposed MEH-FEST detector is

\begin{itemize}
\item
{\it simple}. The method simply focuses on the key information of a given audio and is easy to implement.

\item
{\it fast}. The detector can process an input audio extremely fast, {\it e.g.},
within 3.4 milliseconds in our experiments.

\item
{\it effective}. As shown in the experiments, the MEH-FEST detector is with a false positive rate of 0\%
and false negative rates of 0.053\% and 0\% for detecting FAKEBOB attacks in
GMM and i-vector speaker verification systems, respectively.
\end{itemize}

For a theoretical analysis, we estimate the energy of an audio with the attacker's perturbations
in high frequencies for a short-time period when a speech is absent.
In particular, we show analytically that the MEH-FEST metric is able to enlarge the variance of perturbations by a large factor and
thus make the perturbations much more perceptible.
Moreover, our analysis connects the standard deviation of perturbations with
the detection threshold of MEH-FEST and indicates the theoretical condition
when our proposed method can work well.
The experimental results verify our theoretical analysis.

Inspired by adaptive attacks proposed in \cite{Tramer},
we further consider how attackers can design a white-box countermeasure to avoid the detection of the MEH-FEST method.
Specifically, two different types of adaptive attacks were studied.
One attempts to reduce the perturbation threshold of an attack.
However, through experiments, we found that such an adaptive attack would reduce the attacking power
and is meanwhile vulnerable to the noise-adding defense method proposed in our previous work \cite{Chang}.
The other adaptive attack attempts to avoid perturbing the short-time period signal in an audio
to keep the same minimum energy calculated by the MEH-FEST method.
We then propose a countermeasure that measures the second minimum energy, which can also effectively distinguish between
original audios and adversarial audios, in the similar way as using the minimum energy.
Such a game between attackers and defenders continues when both sides can obtain the implementation details of the other side.
In our experiments, we demonstrate the performance of adaptive FAKEBOB attacks
and the corresponding countermeasures against them.

The remainder of this paper is structured as follows.
Section \ref{sec:background} reviews related background
and presents important observations that lead to our design.
Section \ref{sec:detector} provides the implementation details of the MEH-FEST detector,
whereas Section \ref{sec:theoretical} gives the theoretical analysis of our proposed method.
Next, Section \ref{sec:countermeasures} discusses two possible adaptive adversarial attacks against our MEH-FEST method
and our proposed countermeasures against these adaptive attacks.
Section \ref{sec:evaluations} evaluates through experiments the performance of the MEH-FEST method against FAKEBOB attacks,
as well as our countermeasures against adaptive FAKEBOB attacks.
Finally, Section \ref{sec:conclusions} concludes this paper and discusses the future work.

\section{Background and Observations}
\label{sec:background}

\subsection{Audio Signal}

\begin{figure*}[htb]
\begin{center}
    \mbox{
      \subfigure[Entire original audio]{\includegraphics[width=6cm]{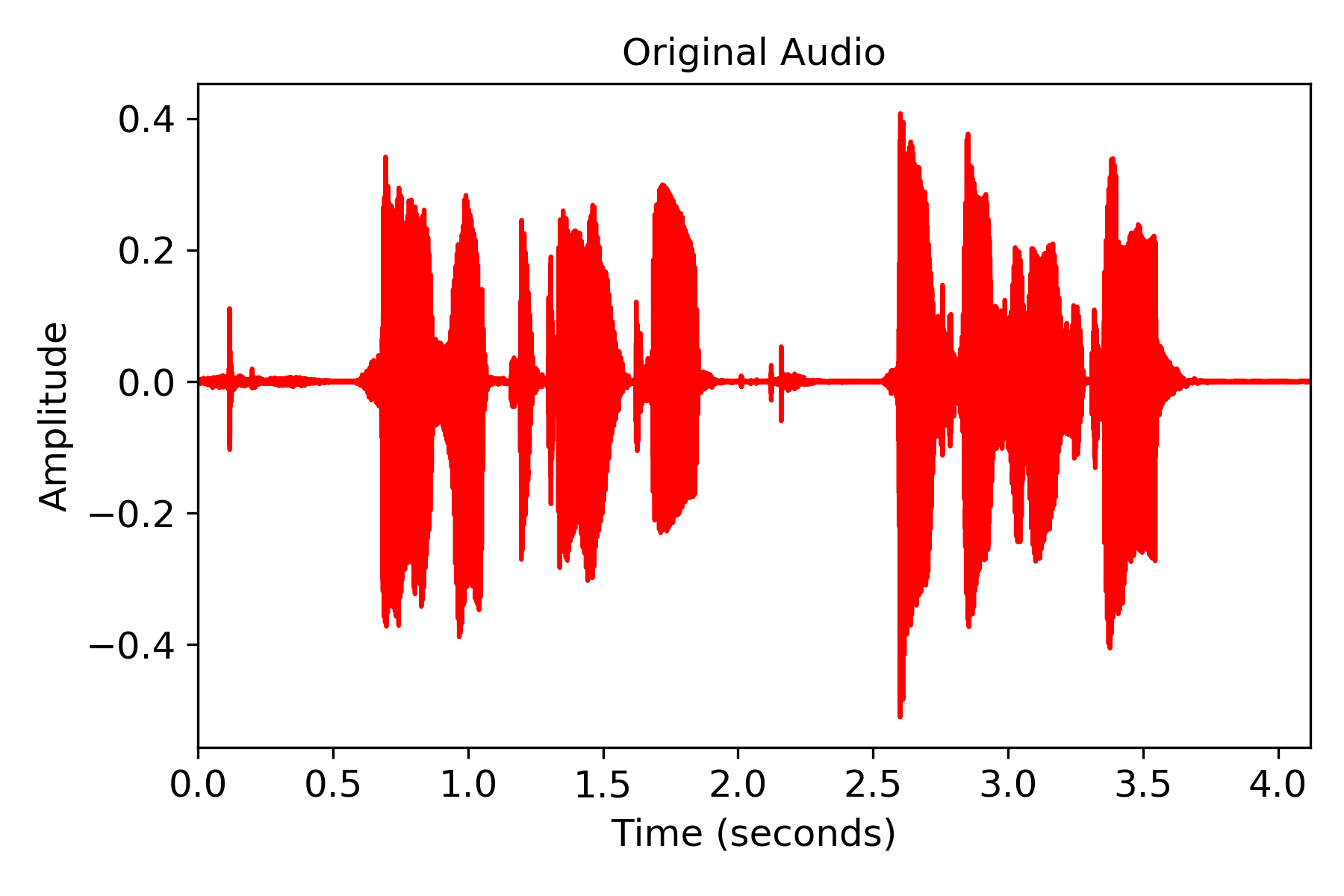}\label{fig:waveform_org}}
      \subfigure[Original audio from 3.5s to 3.532s (when speech is present) ]{\includegraphics[width=6cm]{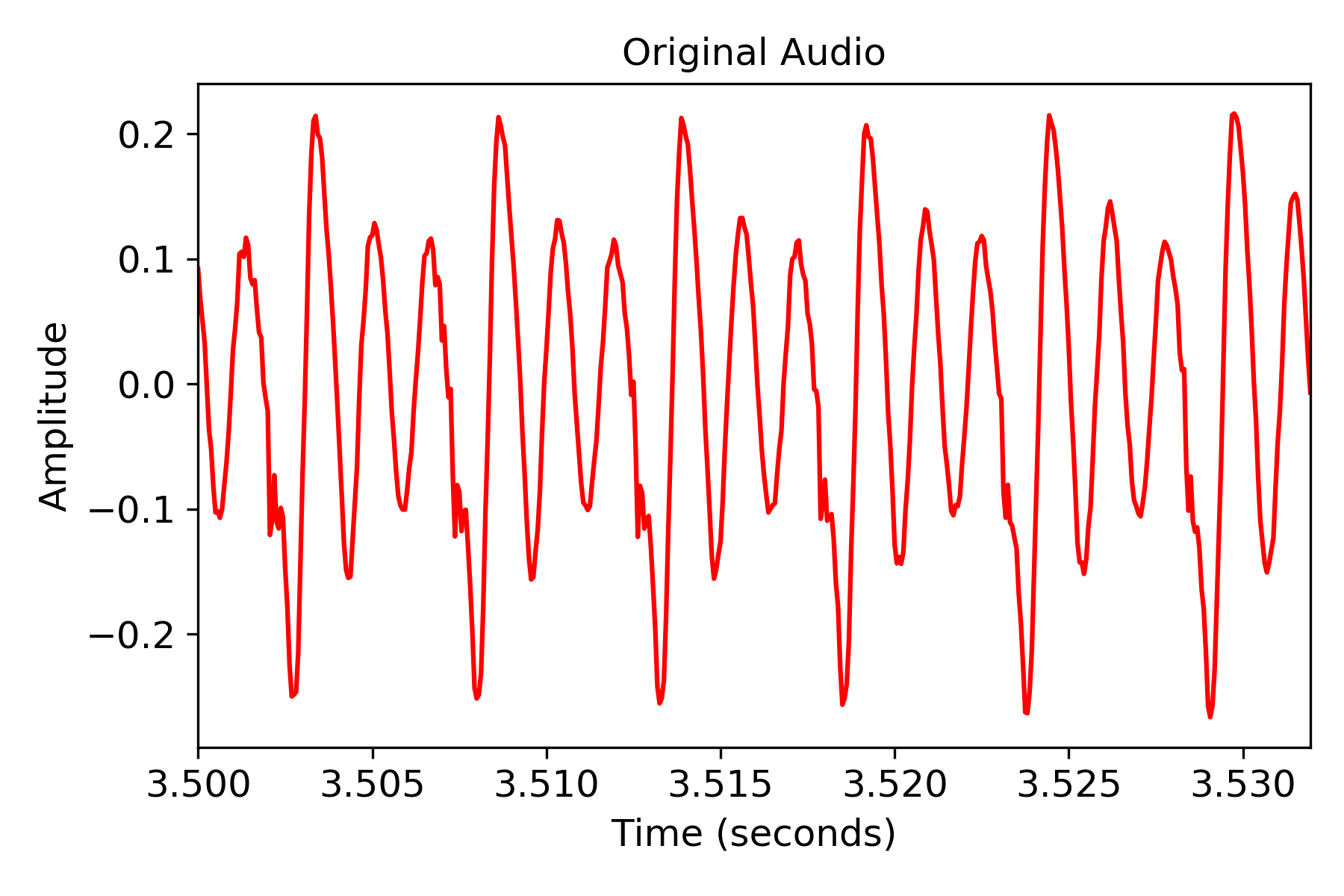}\label{fig:waveform_org_short_time_2}}
      \subfigure[Original audio from 3.98s to 4.012s (when speech is absent) ]{\includegraphics[width=6cm]{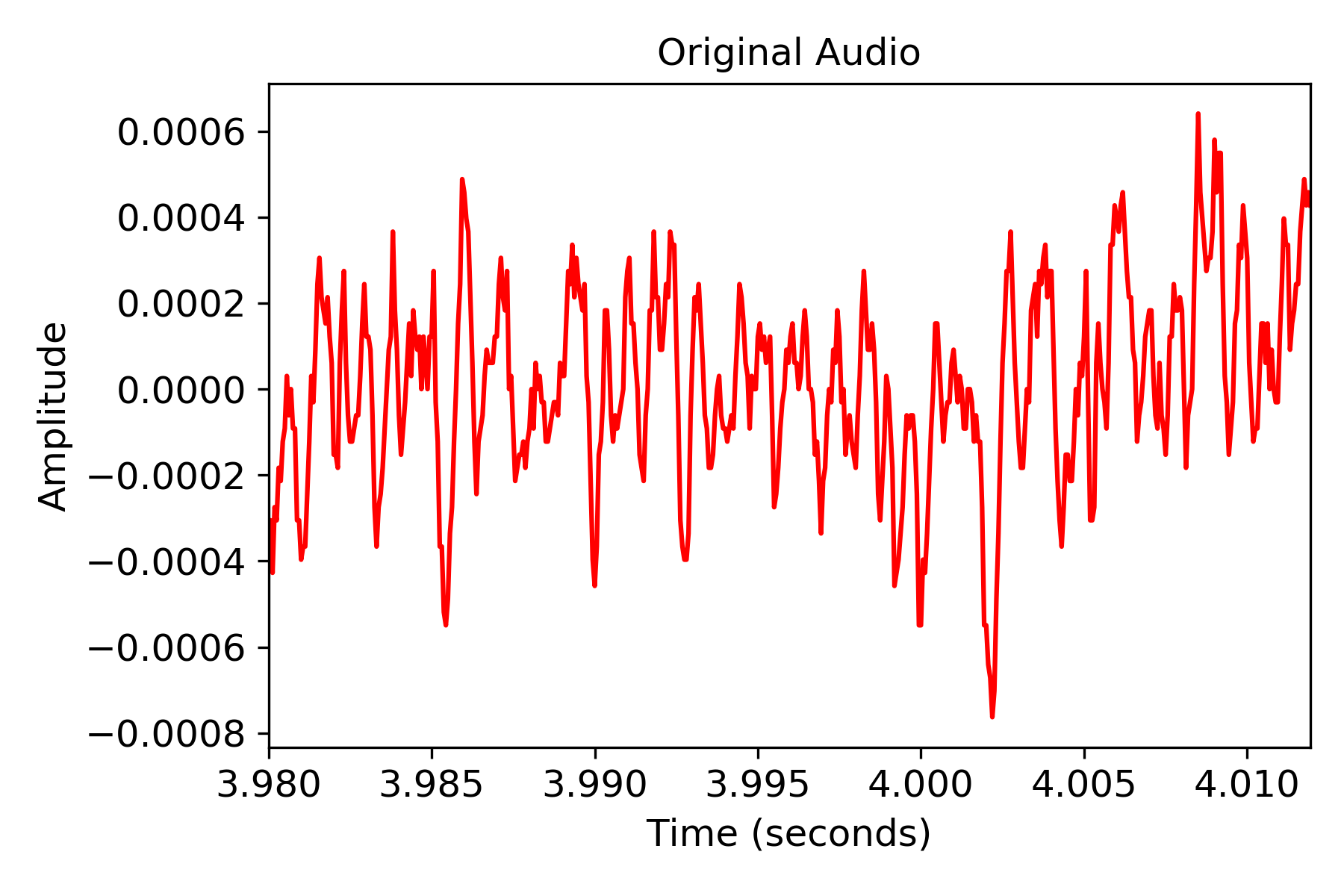}\label{fig:waveform_org_short_time}}
   }
   \caption{Original audio waveform in time domain.}
   \label{fig:org_audio}
   \end{center}
\end{figure*}

It is well known that the hearing frequency range of an audio signal is roughly
between 20 Hz and 20 kHz for humans \cite{Wiki_audio}.
Moreover, the audio signal changes with time.
Figure \ref{fig:waveform_org} plots a waveform of an example audio signal in time domain
and demonstrates how the amplitude of the audio signal varies with time.

One key observation of audio signals is that they are non-stationary \cite{McClellan}.
That is, the statistical properties of the audio signal change with time.
In particular, the audio signal is very different when the speech is present or absent.
For example, Figures \ref{fig:waveform_org_short_time_2} and \ref{fig:waveform_org_short_time} show the same audio signal
in Figure \ref{fig:waveform_org} in two different short-time periods: from 3.5 to 3.532 second and from 3.98 to 4.012 second, respectively.
It can be seen that the amplitude ranges in these two time periods are very different.
The amplitude in Figure \ref{fig:waveform_org_short_time_2} is between -0.27 and 0.22, reflecting the presence of the speech;
but the absolute value of amplitude in Figure \ref{fig:waveform_org_short_time} is very small and less than 0.0008,
indicating the absence of the speech.

The audio signal can be stored in computers in digital format and
is mathematically denoted by $s[n]$, $n = 0, 1, 2, \cdots$ in this paper.
Assuming that the sampling frequency is $f_s$, the relationship between discrete index $n$ and continuous time index $t$
is $t = n / f_s$. For example, when $f_s = 16$ kHz and $t = 3.5$ second, $n = t\times f_s = 56,000$.
In this work, we focus on digital audio signals.

\begin{figure*}[htb]
\begin{center}
    \mbox{
      \subfigure[Entire adversarial audio]{\includegraphics[width=6cm]{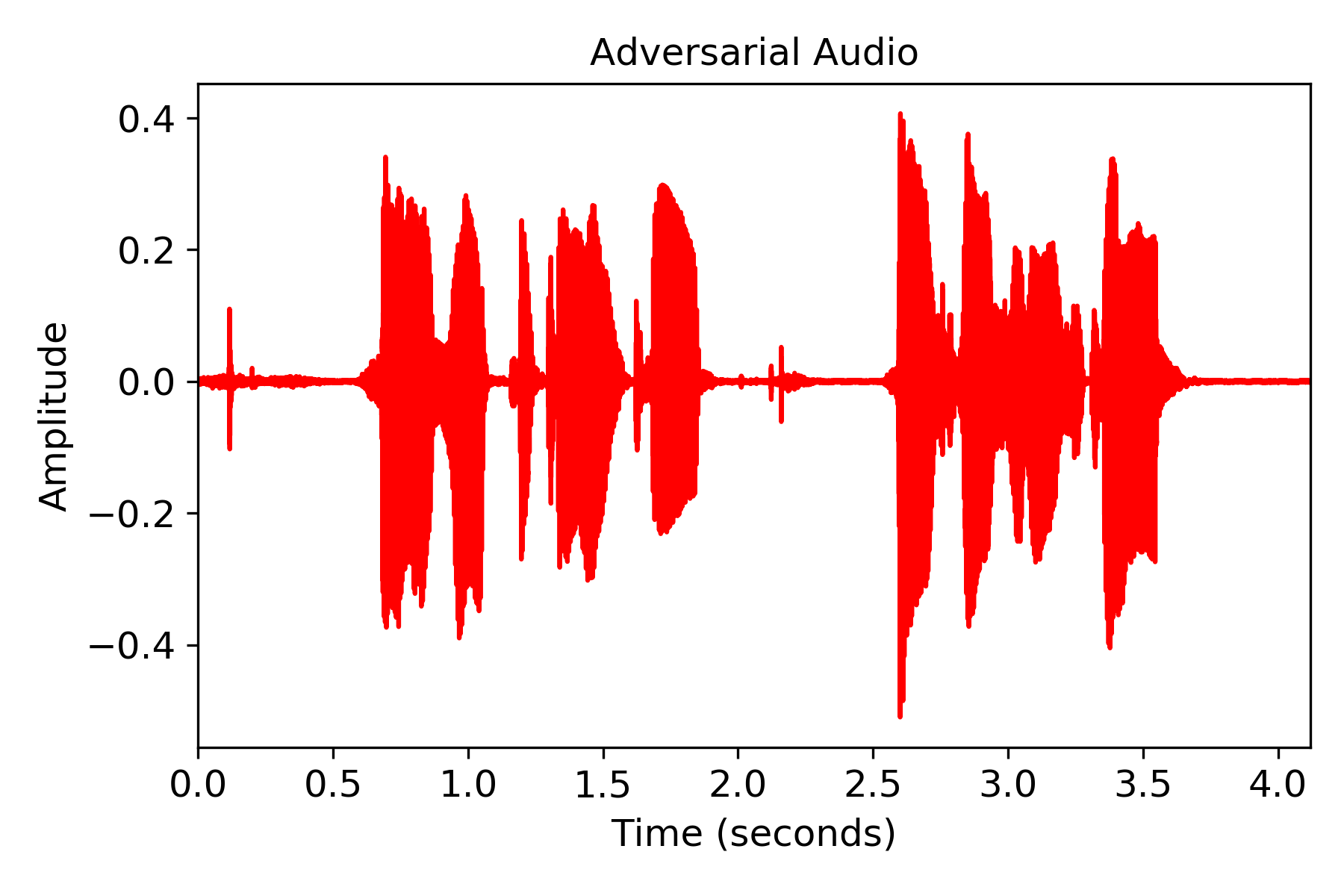}\label{fig:waveform_adv}}
      \subfigure[Adversarial audio from 3.5s to 3.532s (when speech is present) ]{\includegraphics[width=6cm]{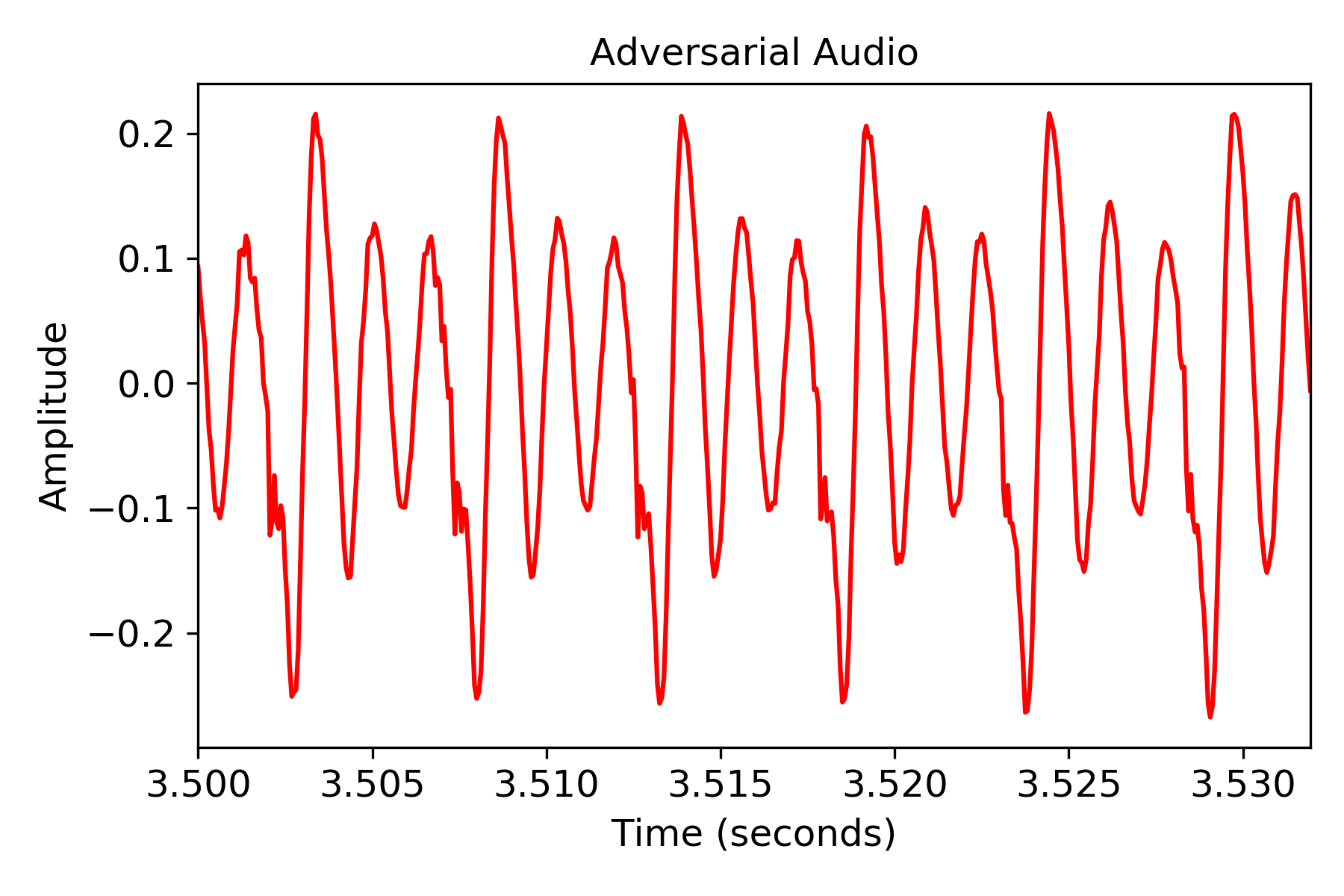}\label{fig:waveform_adv_short_time_2}}
      \subfigure[Adversarial audio from 3.98s to 4.012s (when speech is absent) ]{\includegraphics[width=6cm]{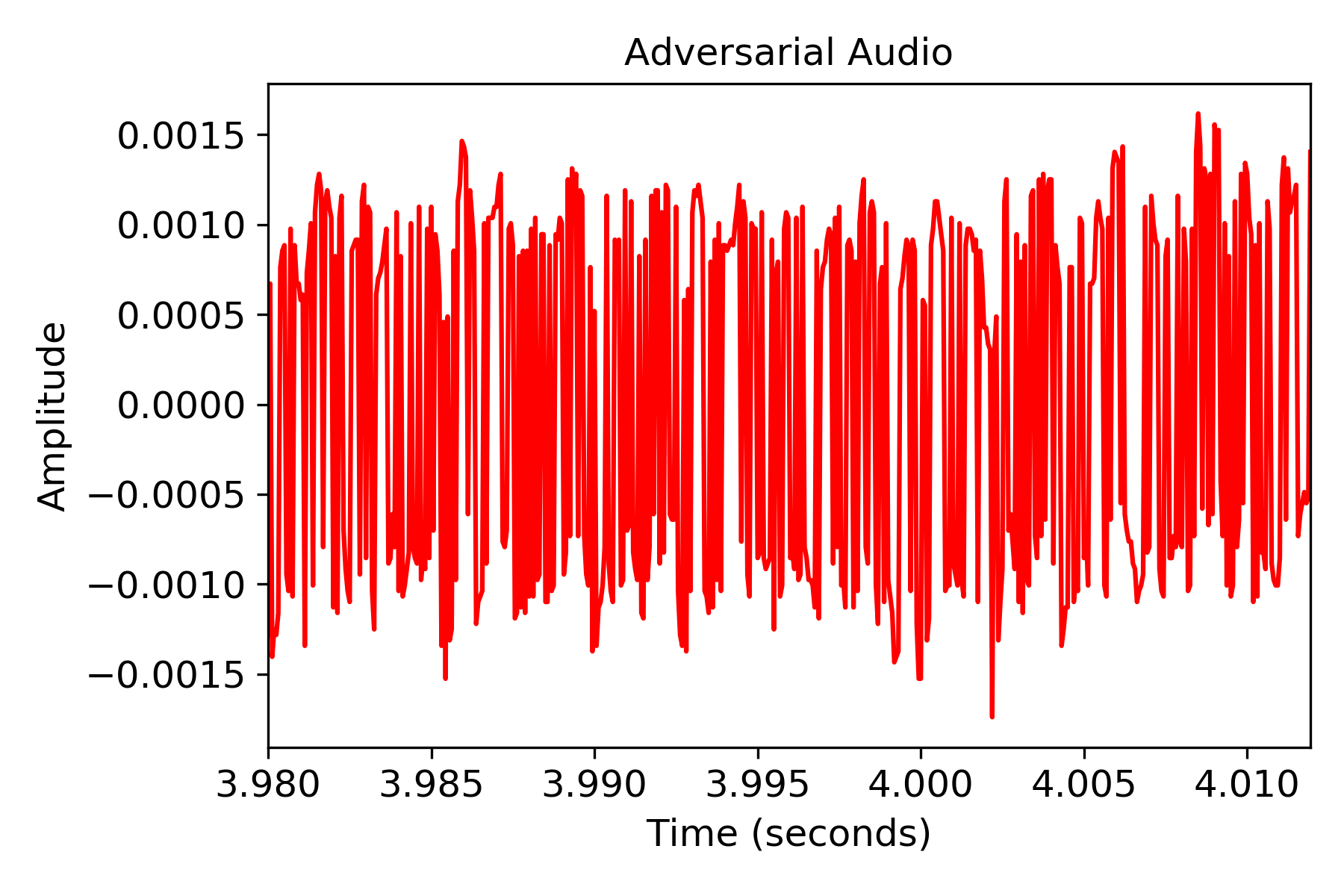}\label{fig:waveform_adv_short_time}}
   }
   \caption{Adversarial audio waveform in time domain.}
   \label{fig:adv_audio}
   \end{center}
\end{figure*}

\subsection{FAKEBOB Attacks Against Speaker Verification Systems}

A speaker verification (SV) system has been applied to determine whether a user is legitimate or illegal.
Currently, the most widely-used SV systems, such as GMM \cite{Reynolds} and i-vector \cite{Dehak}, are score based.
Specifically, the score-based SV system provides a function, $S$, that calculates the score of a given input audio $s[n]$
and then compares the score with a threshold, $\theta$.
If $S(s[n]) \geq \theta$, the SV system would accept $s[n]$; otherwise, it would reject $s[n]$.
There are two main performance metrics for an SV system.
One is the false acceptance rate (FAR), which indicates the percentage of audios from an illegal user
that are falsely accepted by the system.
The other is the false rejection rate (FRR), which presents the percentage of audios from a legitimate user
that are falsely rejected by the system.
The threshold of the SV system, {\it i.e.}, $\theta$, is determined when FAR is equal to FRR,
which is called the equal error rate (EER) \cite{Cheng}.
A smaller EER reflects a better SV system.

An attacker can design an adversarial example attack
to make the SV system falsely accept an illegal user as a legitimate user.
FAKEBOB is the state-of-the-art black-box adversarial attack
against popular score-based SV systems such as GMM and i-vector \cite{Chen}.
The basic idea of FAKEBOB attacks is to find small perturbations $p[n]$, so that an SV system would reject $s[n]$,
but accept $a[n]$, where $s[n]$ is the audio from an illegal user and  $a[n] = s[n] + p[n]$.
Here, we call $s[n]$ as the {\em original audio} and $a[n]$ as the {\em adversarial audio}.
To make the audio imperceptible to humans ({\it i.e.}, adversarial audio $a[n]$ sounds like original audio $s[n]$),
it requires that
$|p[n]| \leq \epsilon$, where $\epsilon$ is called the {\em perturbation threshold} and should be small.
Specifically, FAKEBOB applies the basic iterative method (BIM) \cite{Kurakin} and
the natural evolution strategy (NES) \cite{Ilyas} to find the optimal $p[n]$.
In other words, FAKEBOB attempts to estimate the gradient decent of the objective function over the input audio
to find the direction to change the audio and apply multiple iterations to create an adversarial audio.
The objective function for FAKEBOB attacks is
\begin{equation}
  L(a[n]) = \max \{ \theta - S(a[n]), 0 \},
\end{equation}
and the gradient decent function over the input audio is
\begin{equation}
  f_G(a[n]) = \bigtriangledown_{a[n]} L(a[n]).
\end{equation}
Moreover, FAKEBOB applied a sign function
\begin{equation}
  f_S(x) = \left\{\begin{array} {l l}
1,  & \mbox{if } x > 0  \\
0,  & \mbox{if } x = 0  \\
-1, & \mbox{if } x < 0  \\
\end{array} \right.
\end{equation}
and a clip function
\begin{equation}
  f_C(a[n]) = \left\{\begin{array}{l l}
                      a[n],          & \mbox{if } |a[n]-s[n]| < \epsilon    \\
                      s[n]+\epsilon, & \mbox{if } a[n] \geq s[n] + \epsilon \\
                      s[n]-\epsilon, & \mbox{if } a[n] \leq s[n] - \epsilon \\
                    \end{array} \right.
\end{equation}
Applying these three functions, FAKEBOB updates the input adversarial audio $a[n]$ through the following operation:
\begin{equation}
    a[n] \leftarrow f_C(a[n] - lr \times f_S(f_G(a[n])))
\end{equation}
where $lr$ is the learning rate and can change based on the status of iterations.
The implementation of the FAKEBOB attack is summarized in Algorithm \ref{algo:FAKEBOB}.
In \cite{Chen}, FAKEBOB is shown to be able to achieve a very high targeted attack success rate on popular SV systems.

\begin{algorithm}
\caption{FAKEBOB attack}
\label{algo:FAKEBOB}
\begin{algorithmic}[1]

\STATE {\bf Input:} original illegal audio $s[n]$, threshold of target SV system $\theta$, maximum iteration $I$, score function $S$, clip function $f_C$, learning rate $lr$, sign function $f_S$, and gradient decent function $f_G$

\STATE {\bf Output:} an adversarial audio $a[n]$

\STATE

\STATE $a[n] = s[n]$, for all $n$

\FOR{\texttt{$i = 0$; $i < I$; $i++$ }}

        \IF{$S(a[n]) \geq \theta$}

            \RETURN $a[n]$

        \ENDIF

        \STATE $a[n] = f_C(a[n] - lr \times f_S(f_G(a[n])))$

\ENDFOR

\end{algorithmic}
\end{algorithm}

Using the audio in Figure \ref{fig:waveform_org} as the original audio,
FAKEBOB generated a successful adversarial audio against a GMM SV with $\epsilon = 0.002$,
which is shown in Figure \ref{fig:waveform_adv}.
The waveforms in Figures \ref{fig:waveform_org} and \ref{fig:waveform_adv} are
very similar with incognizable differences to humans.
However, as a result of the non-stationary property of an audio,
the effect of perturbations is significantly different when the speech is present or absent.
We show the waveform of this adversarial audio in two distinct short-time periods,
{\it i.e.}, from 3.5 to 3.532 second and from 3.98 to 4.012 second,
in Figures \ref{fig:waveform_adv_short_time_2} and \ref{fig:waveform_adv_short_time}, respectively.
Comparing Figure \ref{fig:waveform_org_short_time_2} with Figure \ref{fig:waveform_adv_short_time_2},
we can find that when the speech is present, the impact of the perturbations is very minor.
On the other hand, from Figures \ref{fig:waveform_org_short_time} and \ref{fig:waveform_adv_short_time},
it is clear that when the speech is absent,
the perturbations are more significant than the original audio and dominate the signal.
Based on this observation, our designed detector attempts to distinguish between the original audio and the adversarial audio
based on the time period when the speech is absent.

\subsection{Short-Time Fourier Transform}

Short-time Fourier transform (STFT) is a widely used tool for studying audio signals \cite{McClellan,Muller}.
Specifically, an audio signal $s[n]$ can be transformed into the frequency domain by the following equation:
\begin{equation}\label{equ:stft}
  S[k, m] = \sum_{n=0}^{N-1} w[n] s[n+mH] e^{-j2\pi kn/N}, k = 0, 1, \cdots, N/2
\end{equation}
where $w[n]$ is called the analysis window ({\it e.g.}, Hann window) and is used to avoid the ripple artifacts.
The analysis window is with a length of $W$, in which the statistical property of the audio signal does not change much.
$N$ is the length for the fast Fourier transform (FFT) and is assumed to be a power of two \cite{McClellan}.
Note that $N \geq W$.
Moreover, $H$ is called the hop size and is used to specify the step size
in which the window is to be shifted across the signal \cite{Muller}.
Furthermore, $m$ is a non-negative integer and is from 0 to $\lfloor (L-N)/H \rfloor$, where $L$ is the length of the digital audio signal.

\begin{figure*}[htb]
\begin{center}
    \mbox{
      \subfigure[Original audio]{\includegraphics[width=6cm]{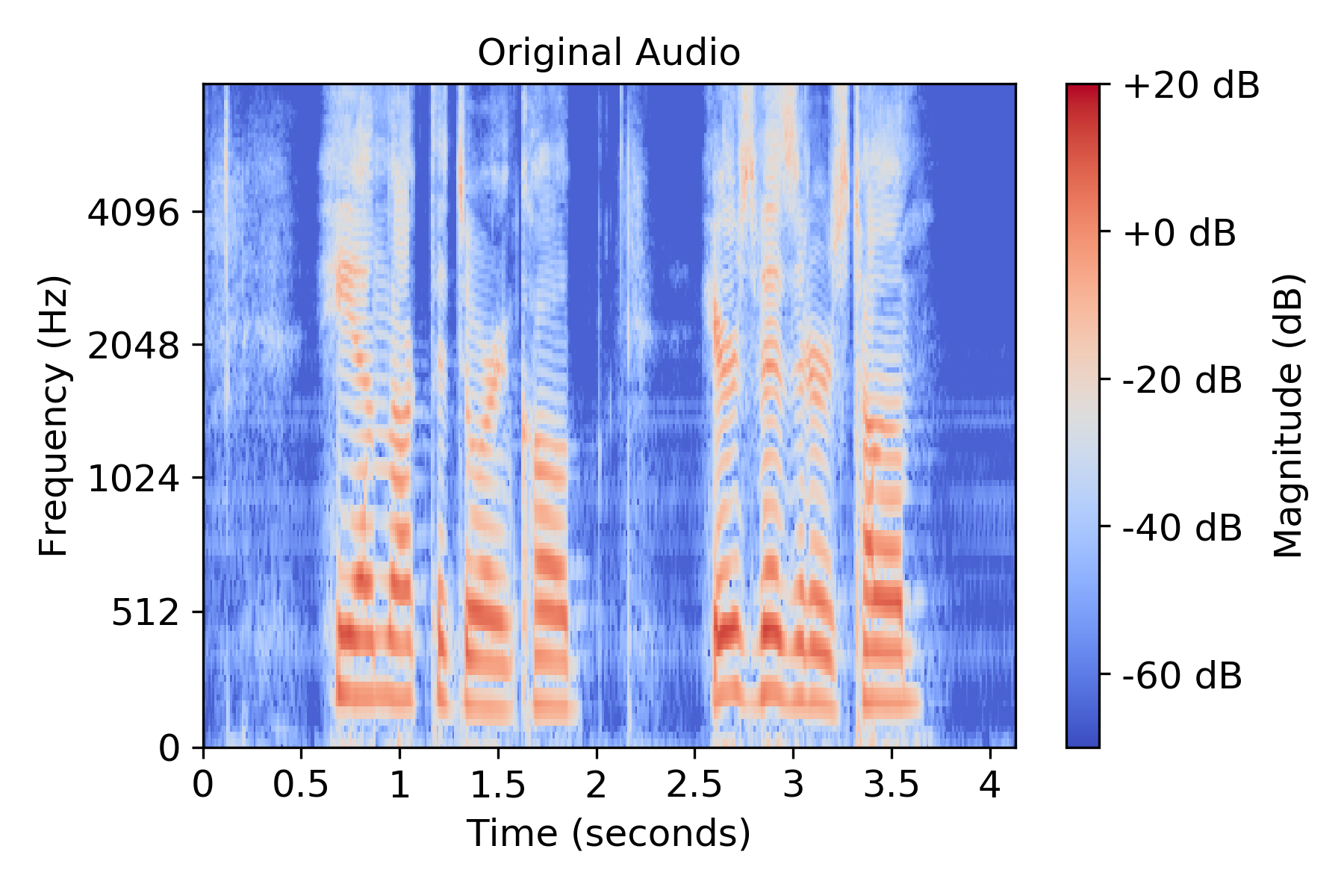}\label{fig:mel_spec_org}}
      \subfigure[Adversarial audio]{\includegraphics[width=6cm]{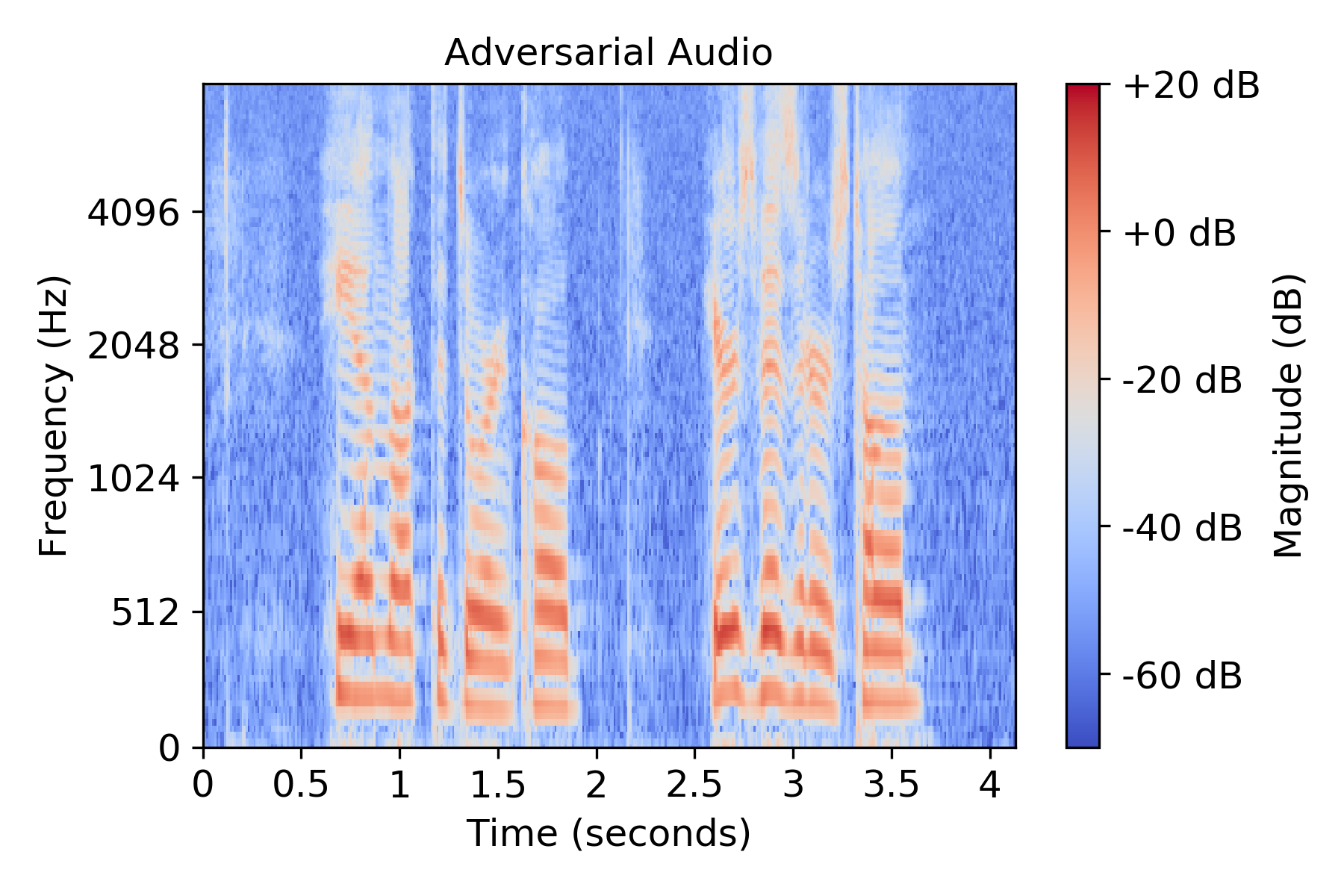}\label{fig:mel_spec_adv}}
      \subfigure[Perturbations]{\includegraphics[width=6cm]{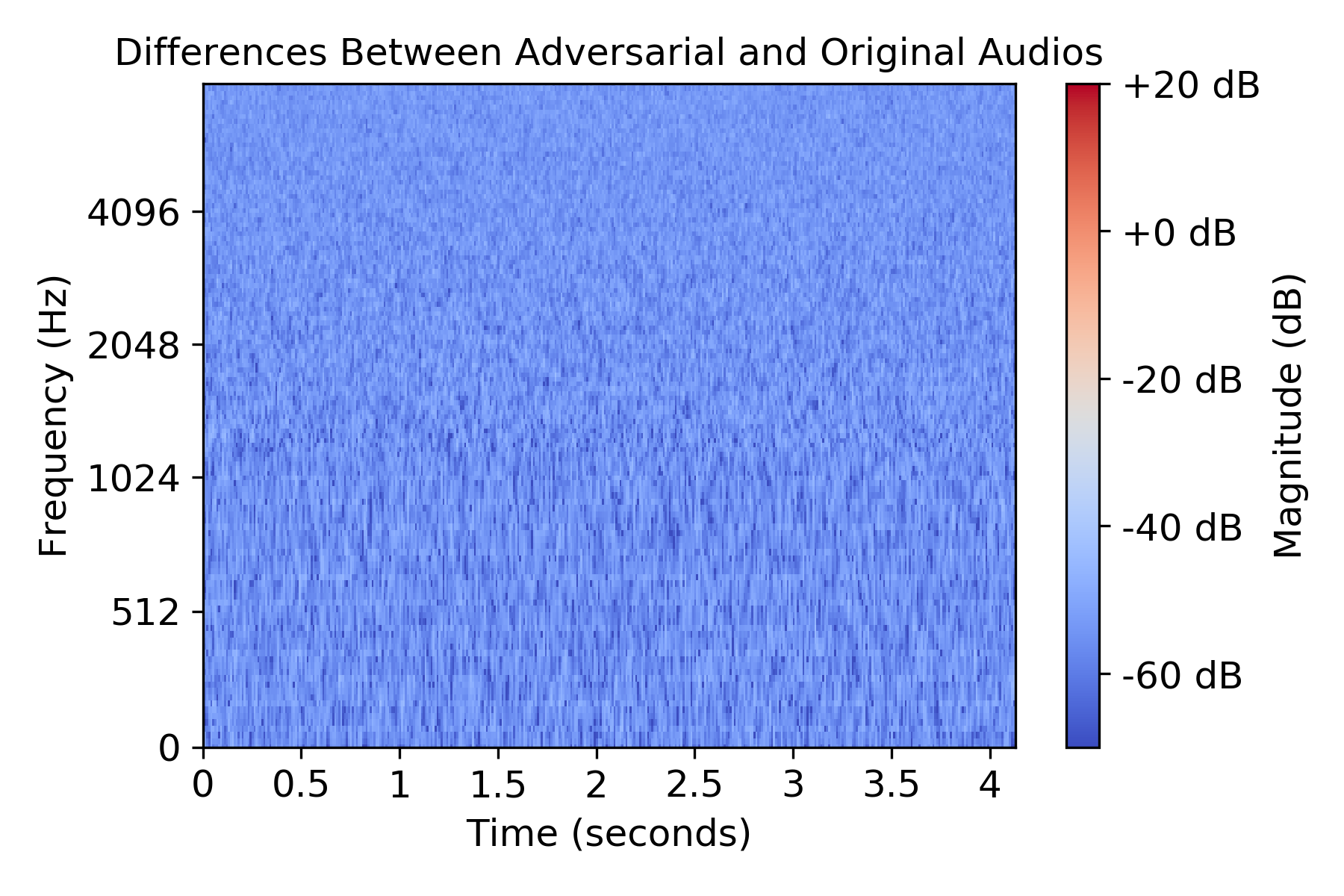}\label{fig:mel_spec_per}}
   }
   \caption{Mel spectrograms of the original audio, the adversarial audio, and the perturbations.}
   \label{fig:mel_spec}
   \end{center}
\end{figure*}

As a result of STFT, $S[k, m]$ contains the information of both time and frequency, where $k$ refers to the frequency and $m$ refers to the time.
A mel spectrogram has been widely used to virtualize the magnitude of the spectrum, {\it i.e.}, $|S[k, m]|$ \cite{McClellan,Muller}.
As shown in Figure \ref{fig:mel_spec}, the x-axis of the mel spectrogram is the time, the y-axis is the frequency in a log scale,
and the color represents the magnitude in dB.
We plot the mel spectrograms of the original audio ({\it i.e.}, $s[n]$ in Figure \ref{fig:waveform_org}) and the adversarial audio ({\it i.e.}, $a[n]$ in Figure \ref{fig:waveform_adv}) in Figures \ref{fig:mel_spec_org} and \ref{fig:mel_spec_adv}, respectively.
The audios are with $f_s = 16$ kHz.
The STFT in these mel spectrograms uses a Hann window with the size $W = 400$ that is equivalent to 25 ms,
the FFT length $N = 512$ that is equivalent to 32 ms, and the hop size $H = 160$ that is equivalent to 10 ms.
It can be seen that although these two mel spectrograms are similar,
the background blue color for the adversarial audio is lighter than that for the original audio,
indicating more energy in the background for the adversarial audio.
We further plot the mel spectrogram of the perturbations ({\it i.e.}, $p[n]$) in Figure \ref{fig:mel_spec_per}.
It is evident that the blue color spreads evenly across time and frequency,
indicating that the perturbations behave in a similar way as white noise.

\begin{figure*}[htb]
\begin{center}
    \mbox{
      \subfigure[Original audio (when speech is absent)]{\includegraphics[width=6cm]{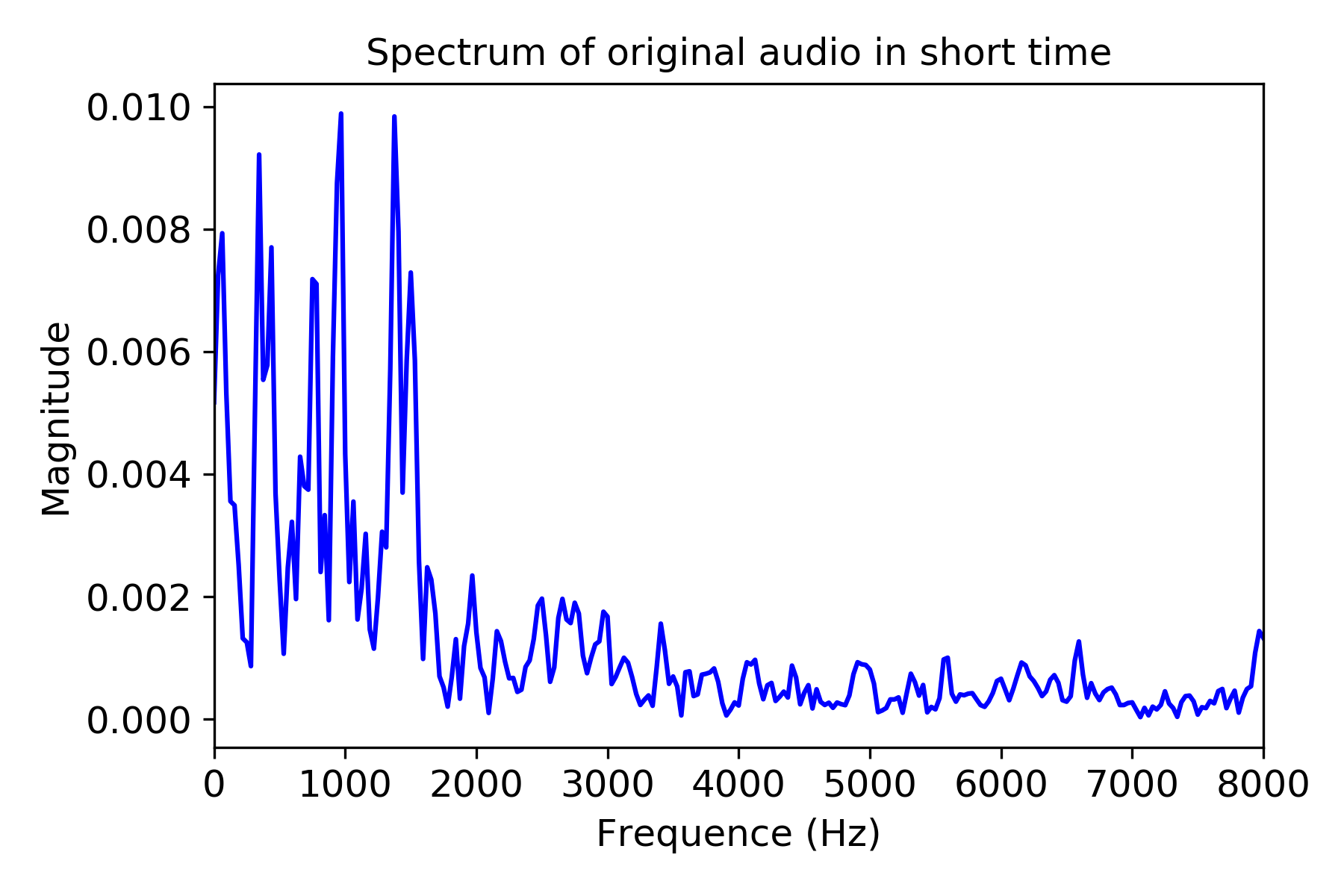}\label{fig:stft_org}}
      \subfigure[Adversarial audio (when speech is absent)]{\includegraphics[width=6cm]{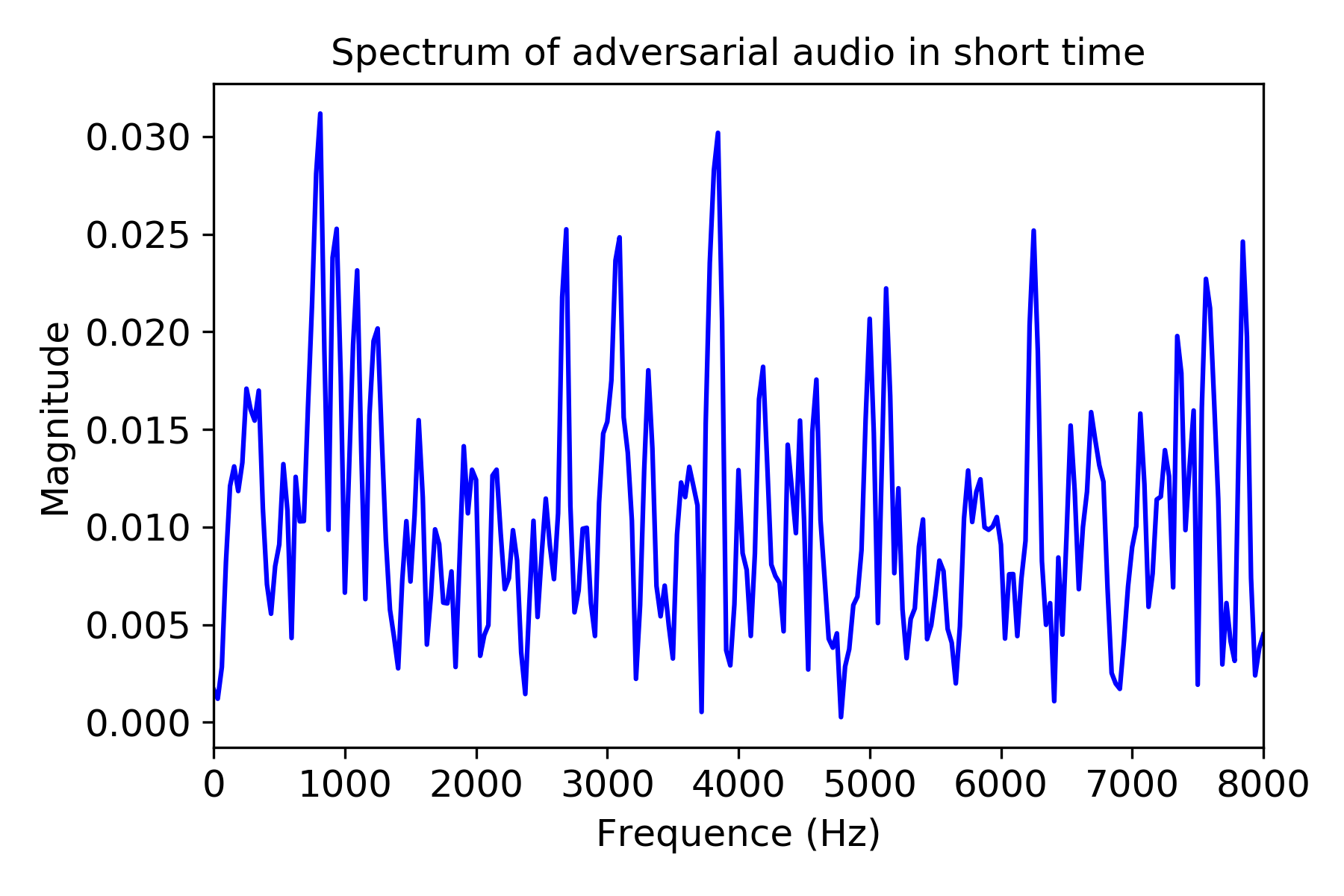}\label{fig:stft_adv}}
      \subfigure[Energy in high frequencies over time frames]{\includegraphics[width=6cm]{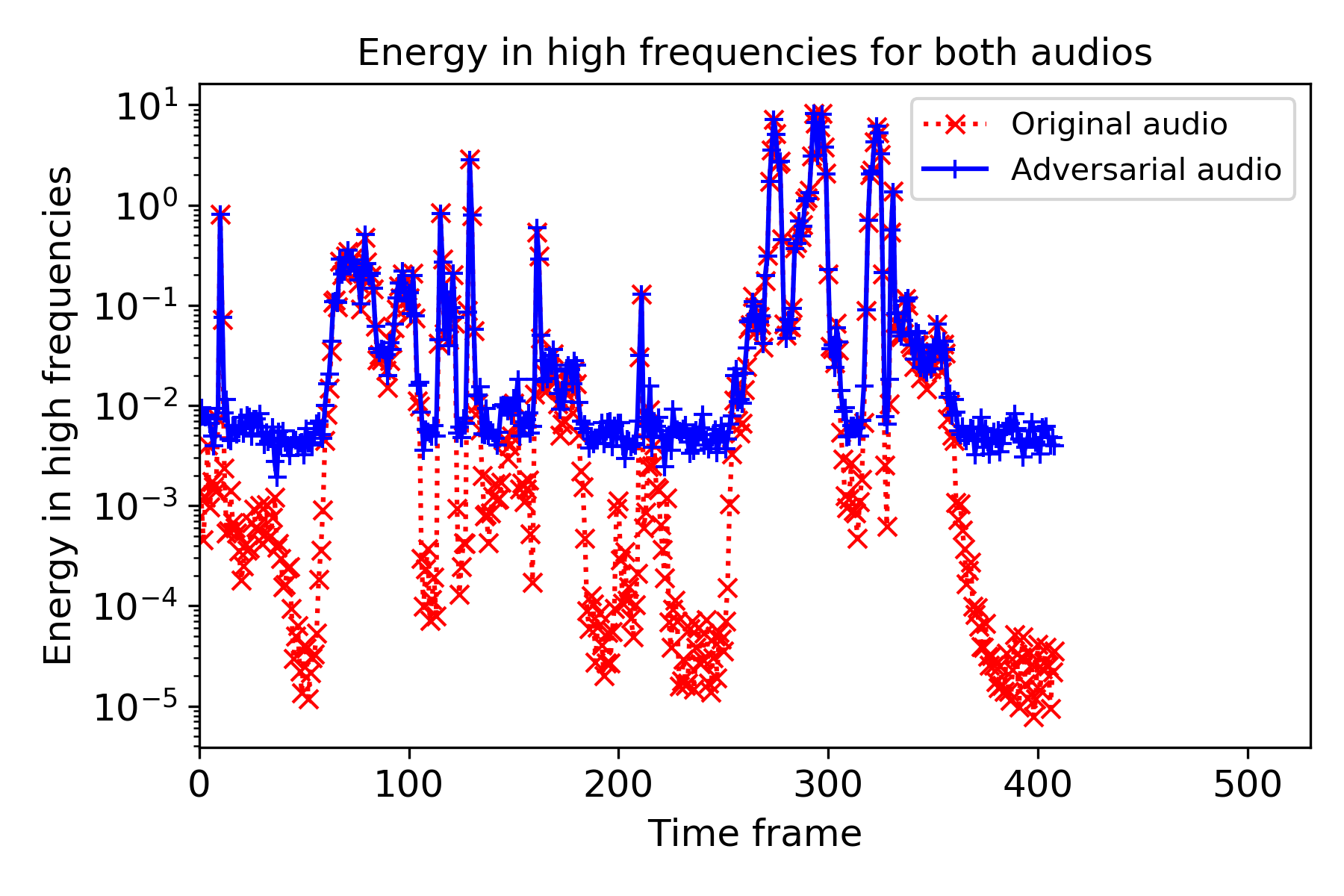}\label{fig:energy_time_frame}}
   }
   \caption{Magnitude of the STFT of the short-time original and adversarial audios, and energy in high frequencies over time frames.}
   \label{fig:stft_energy}
   \end{center}
\end{figure*}

Furthermore, the mel spectrogram of the original audio in Figure \ref{fig:mel_spec_org} indicates that
in general, the magnitude of the signal at higher frequencies is much smaller than that of lower frequency content.
Moreover, comparing the original audio with the adversarial audio in high frequencies,
we observe that the magnitude of the adversarial audio is obviously larger than that of the original audio.
This observation inspires us to focus our detector on high frequencies.

\section{MEH-FEST Detector}
\label{sec:detector}

The MEH-FEST detector attempts to perform a hypothesis testing to decide whether an audio is an original audio
(either from a legitimate user or from an illegal user) or an adversarial audio (either a successful attack
or a failed attack), as shown in the following hypothesis:
\begin{eqnarray}
\nonumber
    \mathcal{H}_0: & &\mbox{the audio is an original audio}    \\
\nonumber
    \mathcal{H}_1: & &\mbox{the audio is an adversarial audio}.
\end{eqnarray}
The main performance metrics to evaluate a detector include the false positive rate
$P_{FP}=P(\mathcal{H}_1|\mathcal{H}_0)$ and the false negative rate
$P_{FN}=P(\mathcal{H}_0|\mathcal{H}_1)$.
The goal of our designed detector is to make both $P_{FP}$ and $P_{FN}$
as small as possible.

To further understand the effect of perturbations when the speech is absent,
we plot the magnitude of the STFT of both the short-time original audio (from Figure \ref{fig:waveform_org_short_time})
and the short-time adversarial audio (from Figure \ref{fig:waveform_adv_short_time})
in Figures \ref{fig:stft_org} and \ref{fig:stft_adv}, respectively.
Since these two short-time audios last only 32 ms that is equal to the time length for $N$,
$m$ is a fixed number in $S[k,m]$, and
we can plot how the magnitude varies with the frequency in Figures \ref{fig:stft_org} and \ref{fig:stft_adv}.
It can be seen that for the original audio, the magnitude is very small when the frequency is high.
On the other hand, the magnitude of the adversarial audio is much larger than that of the original audio
at high frequencies.
Based on this observation, we calculate the energy of the audio signal among high frequencies, {\it i.e.},
\begin{equation}\label{equ:energy_HF}
  E_r[m] = \sum_{k \geq f_t} \left| S[k,m] \right|^2,
\end{equation}
where $f_t$ is the {\em frequency threshold} to determine the range of high frequencies.
For example, if we consider the high frequency range above 7 kHz, $f_t = (N/2) \times (7k / 8k) = 224$,
and thus the energy of the original audio in Figure \ref{fig:stft_org} can be calculated as $E_r[m] = 7.7 \times 10^{-6}$,
whereas for the adversarial audio in Figure \ref{fig:stft_adv}, $E_r[m] = 5.6 \times 10^{-3}$.
Through this example we illustrate that the energy in high frequencies is significantly different for these two audios
when the speech is absent.

How can we find the time frame ({\it i.e.}, $m$) in which the speech is absent?
To identify a proper $m$, we plot how $E_r[m]$ varies with $m$ for both original
and adversarial audios (from Figures \ref{fig:waveform_org} and \ref{fig:waveform_adv}) in Figure \ref{fig:energy_time_frame}.
In this figure, the y-axis uses a log scale to make the differences between two audios more visible.
It can be seen that in many time frames, $E_r[m]$ of the adversarial audio is larger than that of the original audio.
Most importantly, in all time frames, $E_r[m]$ of the adversarial audio is no less than $1.9\times 10^{-3}$,
whereas the minimum of $E_r[m]$ for the original adversarial is only $7.7 \times 10^{-6}$.
This provides a heuristic that when $E_r[m]$ is minimal,
the corresponding $m$ indicates the time frame in which the speech is absent.
Therefore, we have found a metric that can be used for our MEH-FEST detector:
\begin{equation}\label{equ:MEH-FEST}
  E = \min_{m} E_r[m] = \min_{m} \sum_{k \geq f_t} \left| S[k,m] \right|^2.
\end{equation}
Essentially, our detector calculates the minimum energy in high frequencies for the STFT of an audio.

This metric $E$ can be utilized to determine whether an audio is an original audio or an adversarial audio as follows:
\begin{equation}
\label{equ:testing}
\left\{\begin{array} {l l}
\mbox{If}\ E \le D, & \mbox{the audio is an original audio ($\mathcal{H}_0$)}  \\
\mbox{If}\ E >   D, & \mbox{the audio is an adversarial audio ($\mathcal{H}_1$),}  \\
\end{array} \right.
\end{equation}
where $D$ is called the {\em detector threshold} and is a user selected constant.

How can we find a proper value for $D$?
In the perspective of machine learning, we can regard our detection problem as an unsupervised machine learning problem
and estimate $D$ from the existing original audios, either from legitimate users or from illegal users.
Specifically, before applying the MEH-FEST detector to test an audio,
we calculate $E$ in Equation (\ref{equ:MEH-FEST}) for a list of trusted original audios.
We then find the mean value and the standard deviation of these $E$'s, which are denoted as $u_E$ and $\sigma_E$, respectively.
Thus, $D$ can be estimated by the following:
\begin{equation}\label{equ:detection_threshold}
  D = u_E + k \sigma_E,
\end{equation}
where $k$ is a controllable parameter to adjust the detection threshold.
The selection of $k$ affects $P_{FP}$ and $P_{FN}$.
In this work, we choose $k = 3$.

In summary, the MEH-FEST detection method is given in Algorithm \ref{algo:MEH-TEST}.
\begin{algorithm}
\caption{MEH-FEST detection method}
\label{algo:MEH-TEST}
\begin{algorithmic}[1]

\STATE {\bf Input:} input audio $s[n]$, analysis window $w[n]$, window length $W$, FFT length $N$, hop size $H$, frequency threshold $f_t$, and detection threshold $D$

\STATE {\bf Output:} whether the input audio $s[n]$ is an adversarial audio

\STATE

\STATE Calculate the STFT $S[k,m]$ of input audio $s[n]$ based on $w[n]$, $W$, $N$, and $H$ from Equation (\ref{equ:stft})

\STATE $E = \min_{m} \sum_{k \geq f_t} \left| S[k,m] \right|^2$

\IF{$E > D$}

\RETURN \TRUE

\ELSE

\RETURN \FALSE

\ENDIF

\end{algorithmic}
\end{algorithm}

\section{Theoretical Analysis of the MEH-FEST Detector}
\label{sec:theoretical}

In this section, we provide the theoretical analysis of our designed MEH-FEST detector.
Specifically, we quantitatively analyze the effect of perturbations on the energy $E$ in Equation (\ref{equ:MEH-FEST})
when a speech is absent.
We consider two cases: single short-time frame and multiple short-time frames.

\subsection{Single Short-Time Frame}

Based on the observation from Figure \ref{fig:mel_spec_per}, we assume that
adversarial perturbations $p[n]$'s are white noise and are independent and identically distributed (i.i.d.) random variables
that follow a normal distribution
with zero mean and $\sigma^2$ variance, {\it i.e.},
\begin{equation}
    p[n] \sim N(0, \sigma^2).
\end{equation}
That is, $E[p[n]] = 0$, and $E[p^2[n]] = \sigma^2$. Moreover,
\begin{equation}\label{equ:independent_noise}
  E[p[n]p[l]] = 0, \ \mbox{when}\ \ n \neq l.
\end{equation}
Here, the standard deviation $\sigma$ is affected by the perturbation threshold $\epsilon$.
When $\epsilon$ increases, $\sigma$ also increases.
Furthermore, since $|p[n]| \leq \epsilon$, $\sigma^2 = E[p^2[n]] \leq \epsilon^2$, which means $\sigma \leq \epsilon$.

We study the short-time period when a speech is absent in this section.
When a speech is absent, the original audio signal is very small and is assumed to be zero, {\it i.e.}, $s[n] = 0$.
As a result, the adversarial audio $a[n] = s[n]+p[n] = p[n]$, containing only perturbations.

Applying the STFT in Equation (\ref{equ:stft}) to an adversarial audio for a single short-time frame when $s[n] = 0$, we have
\begin{equation}
  S[k] = \sum_{n=0}^{N-1} w[n]p[n] e^{-j2\pi kn/N}, \ k = 0, 1, \cdots, N/2.
\end{equation}
Note that the expectation of $S[k]$ is zero.
Setting $q[n] = w[n]p[n]$, we derive the expectation of $|S[k]|^2$ in the following:
\begin{eqnarray}
\nonumber
   & & E\left[|S[k]|^2\right] \\
\nonumber
   &=& E\left[\left(\sum_{n=0}^{N-1} q[n] e^{-j2\pi kn/N}\right)\left(\sum_{l=0}^{N-1} \overline{q[l]}e^{j2\pi kl/N}\right)\right] \\
   &=& \sum_{n=0}^{N-1} \sum_{l=0}^{N-1} E\left[q[n] \overline{q[l]}\right] e^{-j2\pi k(n-l)/N}
\end{eqnarray}
Since $q[l]$ is a real signal, $\overline{q[l]} = q[l]$.
Moreover, when $n \neq l$, $E[q[n]q[l]] = w[n]w[l]E[p[n]p[l]] = 0$ based on Equation (\ref{equ:independent_noise}).
Therefore,
\begin{eqnarray}
  E\left[|S[k]|^2\right] &=& \sum_{n=0}^{N-1} E[q^2[n]] \\
                         &=& \sigma^2 \sum_{n=0}^{N-1} w^2[n]
\end{eqnarray}

Analysis window $w[n]$ can take different forms.
In this work, we apply the widely-used Hann window \cite{Hann} as the analysis window, {\it i.e.},
\begin{equation}
  w[n] = 0.5 - 0.5\cos\left(\frac{2\pi n}{W-1}\right), \ 0\leq n \leq W-1.
\end{equation}
Using the continuous-time integral as an approximation to the discrete-time summation, we find that
\begin{eqnarray}
\nonumber
  & & \sum_{n=0}^{N-1} w^2[n]  \\
  &\approx& \int_{0}^{W-1} \left[0.5 - 0.5\cos\left(\frac{2\pi x}{W-1}\right)\right]^2 dx \\
  &=& \frac{W-1}{8\pi} \int_{0}^{2\pi} [1-\cos (y)]^2 dy \\
  &=& \frac{3}{8} (W-1)
\end{eqnarray}
Therefore,
\begin{equation} \label{equ:Gamma}
  E\left[|S[k]|^2\right] = \frac{3}{8} (W-1) \sigma^2
\end{equation}
Putting the above equation into Equation (\ref{equ:MEH-FEST}), we have
\begin{equation} \label{equ:theory_E}
  E = \sum_{k \geq f_t} E\left[|S[k]|^2\right] = \frac{3}{8} (W-1) M \sigma^2
\end{equation}
where $M = N/2 - f_t + 1$, indicating the number of discrete frequency content in high frequencies.

Equation (\ref{equ:theory_E}) provides the expected value of the energy in high frequencies for a single short-time frame
when a speech is absent.
In an audio, however, there are multiple short-time frames when a speech is absent or the energy of the speech among high frequencies is very small.
Since $E$ is the minimum value of the multiple energies in these time frames as shown in Equation (\ref{equ:MEH-FEST}),
$E \le 3(W-1)M\sigma^2/8$.

\subsection{Multiple Short-Time Frames}

We consider multiple short-time frames when a speech is absent or the energy of the speech among high frequencies is very small,
{\it i.e.}, $m = 0, 1, 2, \cdots, F-1$, where $F$ is the number of short-time frames considered and $F > 1$.
Since in Equation (\ref{equ:stft}) $s[n + mH] = p[n + mH]$ when a speech is absent or the energy of the speech among high frequencies is very small,
$s[n + mH]$'s are {\em i.i.d.} random variables following a normal distribution with zero mean and $\sigma^2$ variance.
Moreover, in STFT $S[k, m]$'s are a linear combination of $s[n + mH]$'s.
As a result, $S[k, m]$'s are random variables that follow a complex normal distribution \cite{Wiki_complex_normal}
with zero mean and the covariance of $\Gamma=3(W-1)\sigma^2/8$ based on Equation (\ref{equ:Gamma}).
Furthermore, $E_r[m]$ in Equation (\ref{equ:energy_HF}) is the sum of the squares of $M$ normal random variables.
If we set
\begin{equation}
  X_r[m] = \frac{8E_r[m]}{3(W-1)\sigma^2} = E_r[m]/\Gamma,
\end{equation}
then $X_r[m]$'s follow the chi-squared distribution \cite{Wiki_chi_squared} with $M$ degrees of freedom, {\it i.e.,}
\begin{equation}
  X_r[m] \thicksim \chi^2_M.
\end{equation}
Note that
\begin{equation} \label{equ:E_chi}
  E = E[\min_{m} E_r[m]] = \Gamma E[\min_{m} X_r[m]],
\end{equation}
where $m = 0, 1, 2, \cdots, F-1$.
When $F = 1$, $E[\min_{m} X_r[m]] = E[X_r[m]] = M$,
so $E$ in Equation (\ref{equ:E_chi}) is reduced to that in Equation (\ref{equ:theory_E}).

Theoretically, to find the expected value of our MEH-FEST metric for an audio,
we can start with a list of chi-squared distributed random variables and then find
the expectation of the minimum value of variables in this list, {\it i.e.}, $E[\min_{m} X_r[m]]$.
Finally, $\Gamma E[\min_{m} X_r[m]]$ is the theoretical value of the MEH-FEST metric.

Considering the detector threshold $D$ from Equation (\ref{equ:detection_threshold}),
if we set $E$ to be $D$ in Equation (\ref{equ:E_chi}), we then find
\begin{equation} \label{equ:sigma_D}
  \sigma_D = \sqrt{\frac{8D}{3(W-1)E[\min_{m} X_r[m]]}}
\end{equation}
where $\sigma_D$ is the value of $\sigma$ that can lead to the energy of $D$ in high frequencies when a speech is absent.
As a result, when an adversarial audio has perturbations that are with standard deviation $\sigma$ larger than $\sigma_D$,
our MEH-FEST method can detect it correctly with a high probability.
On the other hand, when $\sigma < \sigma_D$, the adversarial audio may be able to confuse the MEH-FEST detector,
which will be further discussed in the next section.

\begin{figure}[tb]
\centerline{\psfig{figure=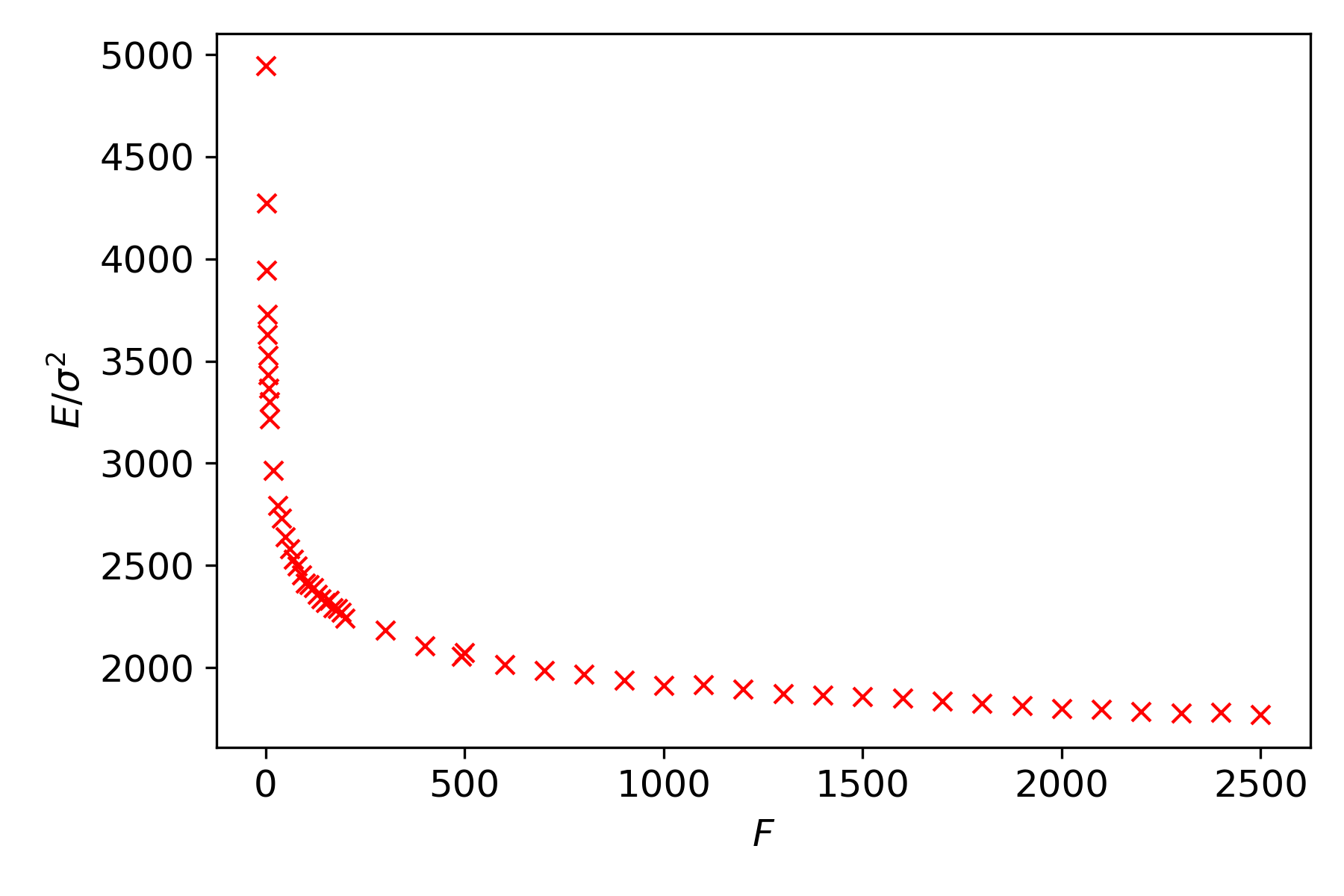, width=7.5cm}}
\caption{Theoretical results for how $E/\sigma^2$ varies with $F$ through the Monte Carlo method.}
\label{fig:theoretical_results}
\end{figure}

We applied the Monte Carlo method \cite{Wiki_monte_carlo} to obtain theoretical $E[\min_{m} X_r[m]]$.
Specifically, in each of 1,000 repeated experiments, $F$ $X_r[m]$'s were obtained by summing the squares of $M$ random variables
that were sampled from a standard normal distribution.
Then, $E[\min_{m} X_r[m]]$ can be approximated by the average of the minimum values of these $F$ $X_r[m]$'s over 1,000 runs.
After getting $E[\min_{m} X_r[m]]$, we can find
\begin{equation} \label{equ:theoretical_result}
  \frac{E}{\sigma^2} = \frac{3}{8}(W-1)E[\min_{m} X_r[m]].
\end{equation}
Figure \ref{fig:theoretical_results} shows how $E/\sigma^2$ varies with $F$ through the Monte Carlo method.
In the experiments, $N = 512$ and $f_t = 224$, so that $M = 33$. Moreover, $W = 400$.
These parameters were also used in our performance evaluation in Section \ref{sec:evaluations}.
It can be seen from Figure \ref{fig:theoretical_results} that when $F$ increases, $E/\sigma^2$ decreases.
Specifically, when $F$ increases from 1 to 2,500, $E/\sigma^2$ decreases from 4,944 to 1,769.
Moreover, when $F \ge 100$, $E/\sigma^2$ has a value around 2,000.
Note that when $F = 492$, $E/\sigma^2 = 2,057$, which we will use to compare with measured results in \ref{sec:valification_analytical}.
From the perspective of treating perturbations as noise, our MEH-FEST metric $E$ is able to enlarge $\sigma^2$ by a factor of 2,057
and make the perturbations much more perceptible.

To estimate the value of $F$, {\it i.e.}, the number of short-time frames, one way is to assume that among high frequencies the energy of a speech is small,
so that the dominating part of $E_r[m]$ is from background noise or perturbations.
As a result, $F$ can be approximated by the number of time frames of STFT in an audio, {\it i.e.}, $\lfloor (L-N)/H \rfloor + 1$.

\section{Attackers' Countermeasures and Defenders' Countermeasures}
\label{sec:countermeasures}

If the implementation detail of the MEH-FEST detector is known to an attacker,
how can this attacker design a white-box countermeasure to avoid the detection?
Such a countermeasure is called an {\em adaptive attack} \cite{Tramer}.
In this section, we study two possible adaptive attacks, as well as the approaches by defenders against these countermeasures.

\subsection{Reducing the Perturbation Threshold $\epsilon$}

As shown in Section \ref{sec:theoretical}, one straightforward countermeasure by attackers is to
reduce the perturbation threshold $\epsilon$
so that the standard deviation of perturbations ({\it i.e.}, $\sigma$)
can be less than $\sigma_D$ in Equation (\ref{equ:sigma_D}).
However, as shown in \cite{Chen} and in our experiments in Section \ref{sec:performance_MEH-FEST},
when $\epsilon$ decreases, the attack success rate of an adversarial attack would decrease as well.
That is, a small value of $\epsilon$ can let adversarial attacks reduce the attacking power in the first place.
Moreover, when $\epsilon$ is small, the resulting adversarial audios may be vulnerable to countermeasures
such as the noise-adding defense system proposed in our previous work \cite{Chang}.
As a result, it is not desirable for attackers to apply a very small value for $\epsilon$.
This will be verified by experimental results in Section \ref{sec:exp_epsilon}.

\subsection{$n$-th FAKEBOB Attacks}

If a FAKEBOB attacker knows that our MEH-FEST detector looks for the short-time period
that leads to the minimum energy $E$ in Equation (\ref{equ:MEH-FEST}),
the attacker can avoid to perturb that specific short-time period during the iteration process.
Specifically, before applying the FAKEBOB attack, an attacker uses a method
similar to the MEH-FEST method in Algorithm \ref{algo:MEH-TEST}
to the original illegal audio, in order to identify the time frame $T$ that leads to $E$, {\it i.e.},
\begin{equation}\label{equ:T}
  T = \argmin_{m} E_r[m] = \argmin_{m} \sum_{k \geq f_t} \left| S[k,m] \right|^2.
\end{equation}
Then, in the execution of the FAKEBOB attack, it avoids to perturb the short-time period between $T\cdot H$ and $T\cdot H + N -1$
in original audio $s[n]$, where $H$ is the hop size and $N$ is the FFT length in Equation (\ref{equ:stft}).
As a result, it is expected that the resulting adversarial audio $a[n]$ would have the same value of $E$ as
that of the original audio $s[n]$. The attacker's countermeasure is summarized in Algorithm \ref{algo:FAKEBOB_countermeasure},
which we also call as the 1st FAKEBOB attack.

\begin{algorithm}
\caption{FAKEBOB attack with a countermeasure against the MEH-FEST detector or 1st FAKEBOB attack}
\label{algo:FAKEBOB_countermeasure}
\begin{algorithmic}[1]

\STATE {\bf Input:} original illegal audio $s[n]$, threshold of target SV system $\theta$, maximum iteration $I$, score function $S$, clip function $f_C$, learning rate $lr$, sign function $f_S$, gradient decent function $f_G$, analysis window $w[n]$, window length $W$, FFT length $N$, hop size $H$, and frequency threshold $f_t$

\STATE {\bf Output:} an adversarial audio $a[n]$

\STATE

\STATE Calculate the STFT $S[k,m]$ of original audio $s[n]$ based on $w[n]$, $W$, $N$, and $H$ from Equation (\ref{equ:stft})

\STATE $T = \argmin_{m} \sum_{k \geq f_t} \left| S[k,m] \right|^2$

\STATE $a[n] = s[n]$, for all $n$

\FOR{\texttt{$i = 0$; $i < I$; $i++$ }}

        \IF{$S(a[n]) \geq \theta$}

            \RETURN $a[n]$

        \ENDIF

        \STATE $a[n] = f_C(a[n] - lr \times f_S(f_G(a[n])))$

        \STATE $a[n] = s[n]$, for $T\cdot H \leq n < T\cdot H + N$

\ENDFOR

\end{algorithmic}
\end{algorithm}

How can a defender counteract such an adaptive attack?
Note that $E_r[m]$ in Equation (\ref{equ:energy_HF}) covers a list of the energy of the audio signal among high frequencies
over time frames $m$, and $E$ is the smallest element in $E_r[m]$.
When an adaptive attack applies Algorithm \ref{algo:FAKEBOB_countermeasure},
$E$ would be the same for both original illegal audio $s[n]$ and adversarial audio $a[n]$.
To defend against such an attack, an idea is to consider the second minimum element in $E_r[m]$, {\it i.e.},
\begin{equation} \label{equ:E_2}
  E_2 = \min_{m} \{e: e\in E_r[m], e > E\}.
\end{equation}
Here we assume that elements in $E_r[m]$ are distinct.
From Figure \ref{fig:energy_time_frame}, it can be seen that for many time frames,
the value of $E_r[m]$ is near the minimum value for either original or adversarial audios.
As a result, we can expect that $E_2$ can be used to distinguish between original audios and adversarial audios.
The countermeasure against the adaptive attack in Algorithm \ref{algo:FAKEBOB_countermeasure}
is summarized in Algorithm \ref{algo:MEH-TEST_countermeasure},
which we also refer to as the 2nd MEH-FEST detection method.
Note that the detection threshold $D_2$ is different from $D$,
but it can be calculated in a similar way as $D$ by applying Equation (\ref{equ:detection_threshold}),
where $u_E$ and $\sigma_E$ are obtained based on $E_2$ values, instead of $E$ values, from a list of training original audios.

\begin{algorithm}
\caption{2nd MEH-FEST detection method against 1st FAKEBOB attack}
\label{algo:MEH-TEST_countermeasure}
\begin{algorithmic}[1]

\STATE {\bf Input:} input audio $s[n]$, analysis window $w[n]$, window length $W$, FFT length $N$, hop size $H$, frequency threshold $f_t$, and detection threshold $D_2$

\STATE {\bf Output:} whether the input audio $s[n]$ is an adversarial audio

\STATE

\STATE Calculate the STFT $S[k,m]$ of input audio $s[n]$ based on $w[n]$, $W$, $N$, and $H$ from Equation (\ref{equ:stft})

\STATE $E = \min_{m} \sum_{k \geq f_t} \left| S[k,m] \right|^2$

\STATE $E_2 = \min_{m} \{e: e\in \sum_{k \geq f_t} \left| S[k,m] \right|^2, e > E \}$

\IF{$E_2 > D_2$}

\RETURN \TRUE

\ELSE

\RETURN \FALSE

\ENDIF

\end{algorithmic}
\end{algorithm}

The game between attackers and defenders can continue.
If the implementation of both MEH-FEST in Algorithm \ref{algo:MEH-TEST}
and its extension in Algorithm \ref{algo:MEH-TEST_countermeasure} is known to attackers,
they would avoid perturbing two short-time frames that lead to $E$ and $E_2$.
Specifically, we define
\begin{equation}
    T_2 = \argmin_{m}\{ e: e\in E_r[m], e > E \}.
\end{equation}
Then, the attackers generate FAKEBOB adversarial audios by avoiding changing the audio signal in
both $[T\cdot H, T\cdot H + N)$ and $[T_2 \cdot H, T_2 \cdot H + N)$.
In such a way, both $E$ and $E_2$ in an adversarial audio would be the same as the original audio.

As a countermeasure by defenders, they would look into the third smallest element in $E_r[m]$, {\it i.e.}, $E_3$,
as the detection target.
However, attackers can design an adaptive attack that avoid all three time frames that lead to $E$, $E_2$, $E_3$.
We name such a detection method by defenders as $n$-th MEH-FEST,
where $n$ refers to the $n$-th smallest element in $E_r[m]$ and the detection target is $E_n$, {\it i.e.},
\begin{equation}
  E_n = \min_{m} \{e: e\in E_r[m], e > E_{n-1} \}, \ n = 2, 3, 4, \cdots
\end{equation}
and $E_1 = E$.
Similarly, we name the corresponding adaptive attack as $n$-th FAKEBOB, where
\begin{equation}
  T_n = \argmin_{m}  \{e: e\in E_r[m], e > E_{n-1} \}, \ n= 2, 3, 4, \cdots
\end{equation}
and $T_1 = T$.
It is noted that 1st MEH-FEST is the original MEH-FEST method proposed in Algorithm \ref{algo:MEH-TEST},
whereas 1st FAKEBOB is the FAKEBOB attack with the countermeasure
against 1st MEH-FEST shown in Algorithm \ref{algo:FAKEBOB_countermeasure}.

We expect that as $n$ increases, in general the $n$-th FAKEBOB attack would reduce the attack success rate and
increase the running time to generate adversarial audios, because more short-time frames are unchanged from the original audios.
On the other hand, we also expect that the efficiency of the $n$-th MEH-FEST detection method would be reduced
as $n$ increases,
because the values of $E_n$ become closer for original and adversarial audios as $n$ grows.

Note that when calculating the STFT of an audio, {\it i.e.}, $S[k,m]$, there is signal overlapping between two neighboring
time frames when $H < N$.
That is, $S[k, m]$ and $S[k, m+1]$ are calculated with some common $s[n]$'s.
Therefore, when $T$ and $T_2$ are neighboring time frames,
the efficiency of applying $E_2$ in Equation (\ref{equ:E_2})
through Algorithm \ref{algo:MEH-TEST_countermeasure} would be negatively affected against the 1st FAKEBOB attack.
To avoid such an effect, we, as a defender, introduce a constraint that
\begin{equation} \label{equ:constrain}
  |T_2 - T| \geq \lceil N/H \rceil
\end{equation}
in Algorithm \ref{algo:MEH-TEST_countermeasure}.
That is, we would keep searching for the second smallest element in $E_r[m]$
only for those time frames that are at least $\lceil N/H \rceil$ distance from $T$.
Moreover, similar constraints can be applied to $n$-th MEH-FEST.
For example, for 3rd MEH-FEST, besides Inequality (\ref{equ:constrain}),
the following constraints should also be followed: $|T_3-T| \geq \lceil N/H \rceil$,
and $|T_3-T_2| \geq \lceil N/H \rceil$.

\section{Performance Evaluations}
\label{sec:evaluations}

In this section, we first describe the experimental setup.
We then verify the analytical results of our designed MEH-FEST method through experiments.
Next, we evaluate the performance of the MEH-FEST detector against FAKEBOB attacks.
Finally, we show the performance of defenders' countermeasures against attackers' adaptive attacks.

\subsection{Experimental Setup}

We used a virtual machine (VM) in Google Cloud Platform \cite{GCP} to run all our experiments.
The VM is with 16 cores, 64 GB memory, and 3.10 GHz CPU ({\it i.e.}, c2-standard-16 machine type) and is installed with Ubuntu 20.04.
Moreover, we applied the code and the dataset provided in \cite{Chen} to run FAKEBOB attacks against
both GMM and i-vector SV systems, which were implemented by the Kaldi speech recognition toolkit \cite{Povey}.
Specifically, the dataset comes from LibriSpeech \cite{Panayotov} and contains the audios of five legitimate users and four illegal users.
All audios are with a sampling frequency of 16 kHz.
There are 25 audios for each illegal user and a total of 500 audios from all legitimate users.
It is noted that among these 600 original audios, some audios contain perceptible background noise.
We applied these audios in our experiments to see if our proposed MEH-FEST method can distinguish
between normal background noise and malicious perturbations by adversarial attacks.

In our MEH-FEST detector shown in Algorithm \ref{algo:MEH-TEST}, we chose the following parameters for the STFT:
Hann window with a window length $W = 400$ ({\it i.e.}, 25 ms), FFT length $N = 512$ ({\it i.e.}, 32 ms),
and hop size $H = 160$ ({\it i.e.}, 10 ms).
These parameters have been widely applied to calculate the STFT of an audio signal \cite{Muller,Lee}.
We applied the Librosa library \cite{Librosa} to implement the STFT.
Moreover, we used 7 kHz as the frequency threshold, {\it i.e.}, $f_t = 224$,
so that high frequencies are between 7 kHz and 8 kHz.

\subsection{Verification of Analytical Results}
\label{sec:valification_analytical}

\begin{figure}[tb]
\centerline{\psfig{figure=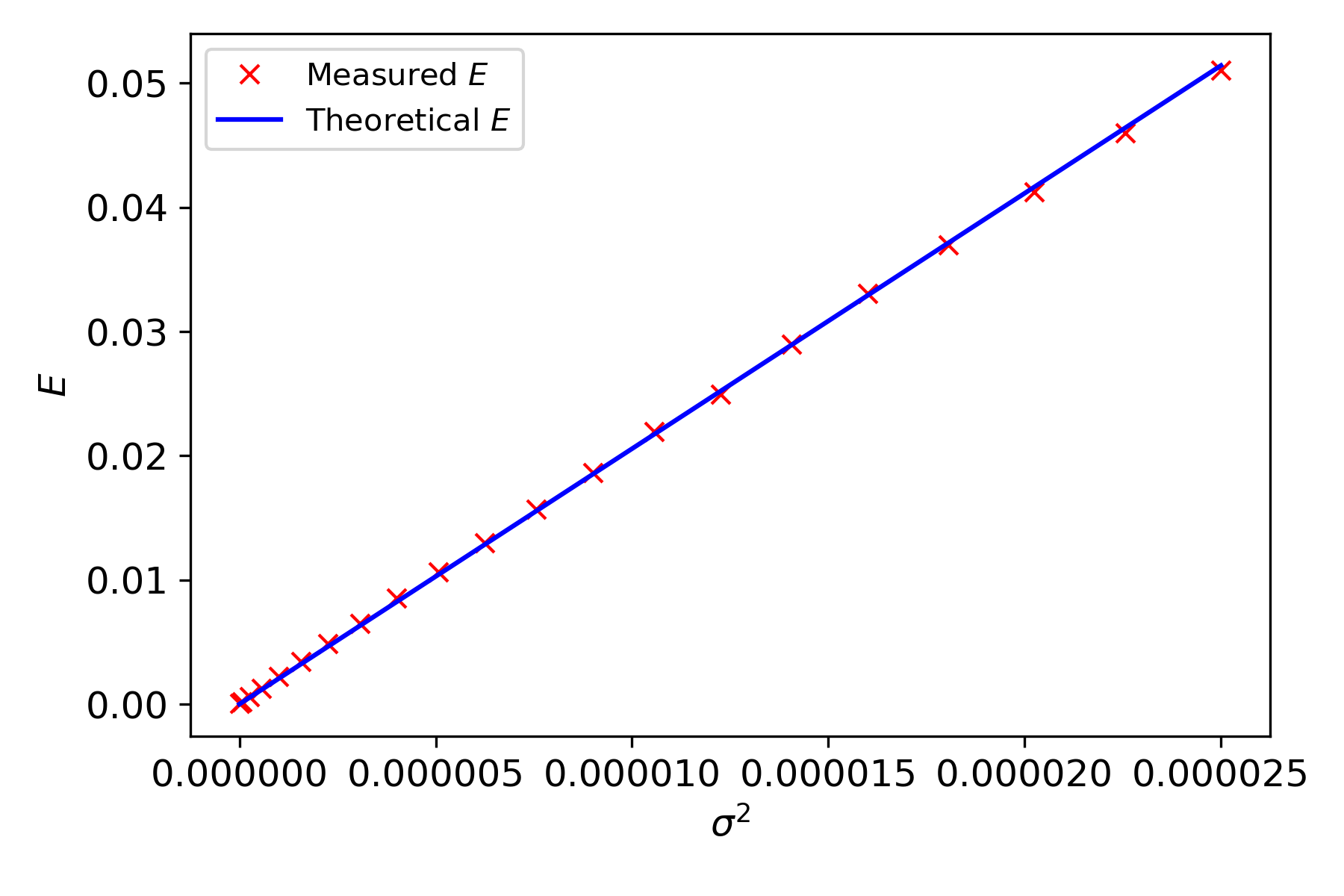, width=7.5cm}}
\caption{Comparing the measured $E$ with the theoretical $E$ for 600 audios with white noise.}
\label{fig:noisy_audios}
\end{figure}

To verify the analytical results of the MEH-FEST detector provided in Section \ref{sec:theoretical},
we added white noise with zero mean and $\sigma^2$ variance to 600 original audios.
We calculated the STFT $S[k, m]$ of an audio with noise based on Equation (\ref{equ:stft}) through the Librosa library
and then measured $E$ based on Equation (\ref{equ:MEH-FEST}).
Figure \ref{fig:noisy_audios} shows how the average of the measured $E$'s over 600 audios varies with $\sigma^2$,
when $\sigma$ increases from 0 to 0.005.

It is noted that the average length of these 600 audios is 4.95 seconds,
which correspond to an average number of samples of 79,219.61, {\it i.e.}, $L = 79,219.61$.
Therefore, we can estimate $F = \lfloor (L-N)/H \rfloor + 1 = 492$.
As shown in Equation (\ref{equ:theoretical_result}) and Figure \ref{fig:theoretical_results},
the corresponding theoretical value of $E/\sigma^2$ when $F = 492$ is 2,057.
It can be seen from Figure \ref{fig:noisy_audios} that the theoretical $E$ accurately predicts the value of the measured $E$.

\subsection{Performance of the MEH-FEST Detector Against FAKEBOB Attacks}
\label{sec:performance_MEH-FEST}

In the FAKEBOB attack, we used 1,000 for the maximum iteration ({\it i.e.}, $I$) as suggested in \cite{Chen}.
Different from the experiments in \cite{Chen}, we registered each legitimate user in a stand-alone SV system
and obtained five different SV systems,
instead of registering all five legitimate users into the same SV system.
As a result, the performance of FAKEBOB attacks is different from that presented in \cite{Chen}.
However, we think that such a setup is more realistic.
It is noted that the proposed value of the perturbation threshold ({\it i.e.}, $\epsilon$) in \cite{Chen} is 0.002.

In a GMM SV system, we implemented the FAKEBOB attacks using 100 original audios from illegal users
and with different values of $\epsilon$.
Table \ref{tab:FAKEBOB} summarizes the performance of FAKEBOB attacks against GMM SV systems.
It can be seen that when $\epsilon$ decreases from 0.005 to 0.0005, the average of the attack success rate (ASR)
over five legitimate users decreases from 97.87\% to 61.47\%,
whereas the total running time increases from 17 hours 32 minutes to 99 hours 5 minutes.
Table \ref{tab:FAKEBOB} also shows the average standard deviation ({\it i.e.}, $\sigma$)
of perturbations in the short-time period $T$ that leads to $E$ in adversarial audios.
It can be seen that $\sigma < \epsilon$.
Moreover, when $\epsilon$ decreases from 0.005 to 0.0005, $\sigma$ decreases from $36.60\times 10^{-4}$ to $3.78\times 10^{-4}$.
In our experiments, the average equal error rate (EER) of five GMM SV systems is 6.20\%.

\begin{table}[htb]
\begin{center}
\caption{FAKEBOB attacks against GMM SV systems.}
\label{tab:FAKEBOB}
\begin{tabular}{|r|c|c|c|c|} \hline
   Perturbation threshold $\epsilon$            & 0.005   & 0.002   & 0.001   & 0.0005       \\ \hline
   Average ASR                                  & 97.87\% & 90.24\% & 77.90\% & 61.47\%      \\
   Total running time                           & 17h 32m & 31h 3m  & 55h 23m & 99h 5m       \\
   Average $\sigma$ in $T$ ($\times 10^4$)      & 36.60   & 15.54   & 7.86    & 3.78         \\ \hline
\end{tabular}
\end{center}
\end{table}

We also run FAKEBOB attacks against i-vector SV systems with $\epsilon = 0.002$.
The average ASR is 95.28\%, the total running time is 441 hours 19 minutes,
and the average $\sigma$ in $T$ of adversarial audios is $14.59\times 10^{-4}$.
Moreover, the average EER of five i-vector SV systems is 2.64\%.

\begin{figure*}[htb]
\begin{center}
    \mbox{
      \subfigure[Original audios]{\includegraphics[width=6cm]{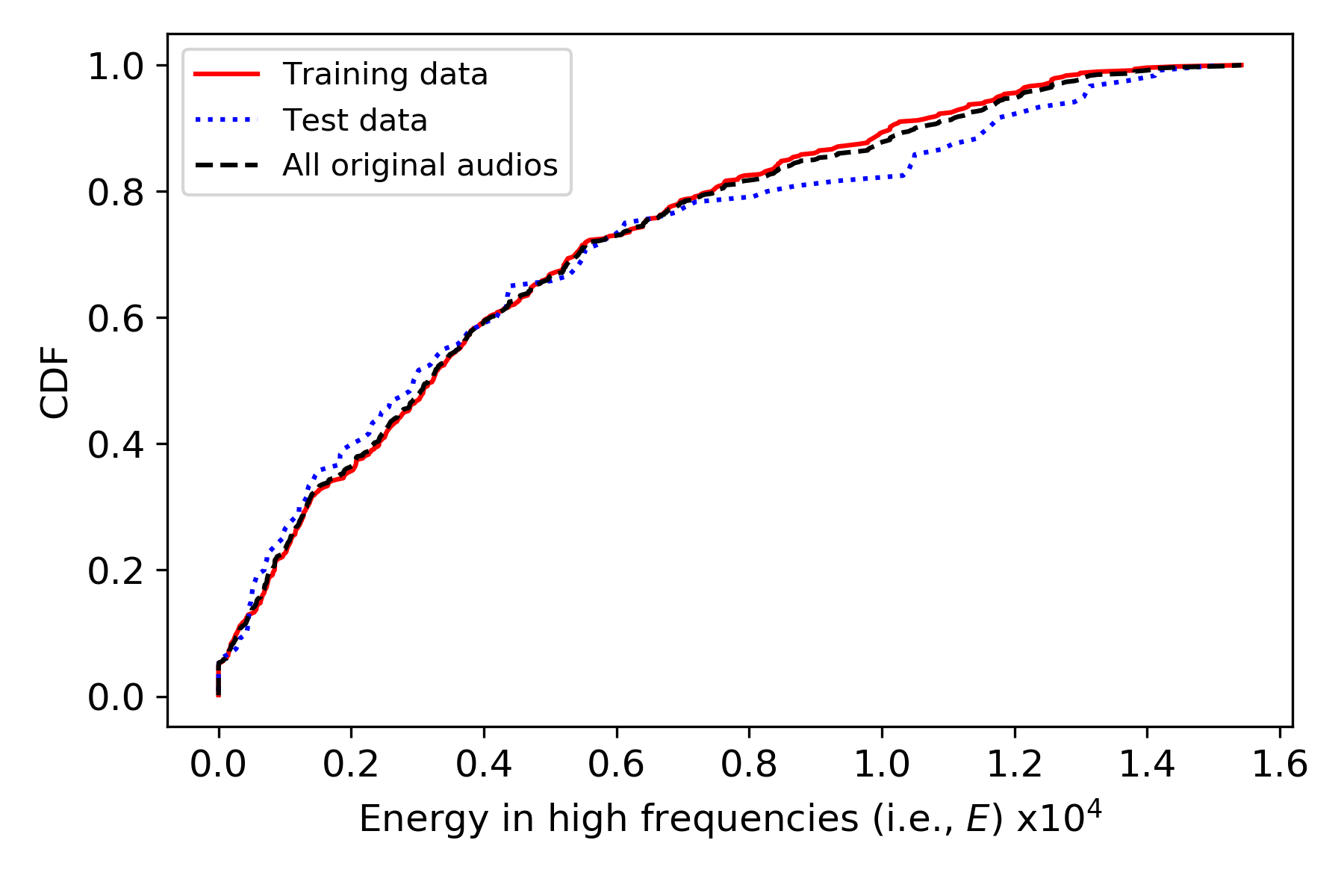}\label{fig:performance_org}}
      \subfigure[GMM SV systems]{\includegraphics[width=6cm]{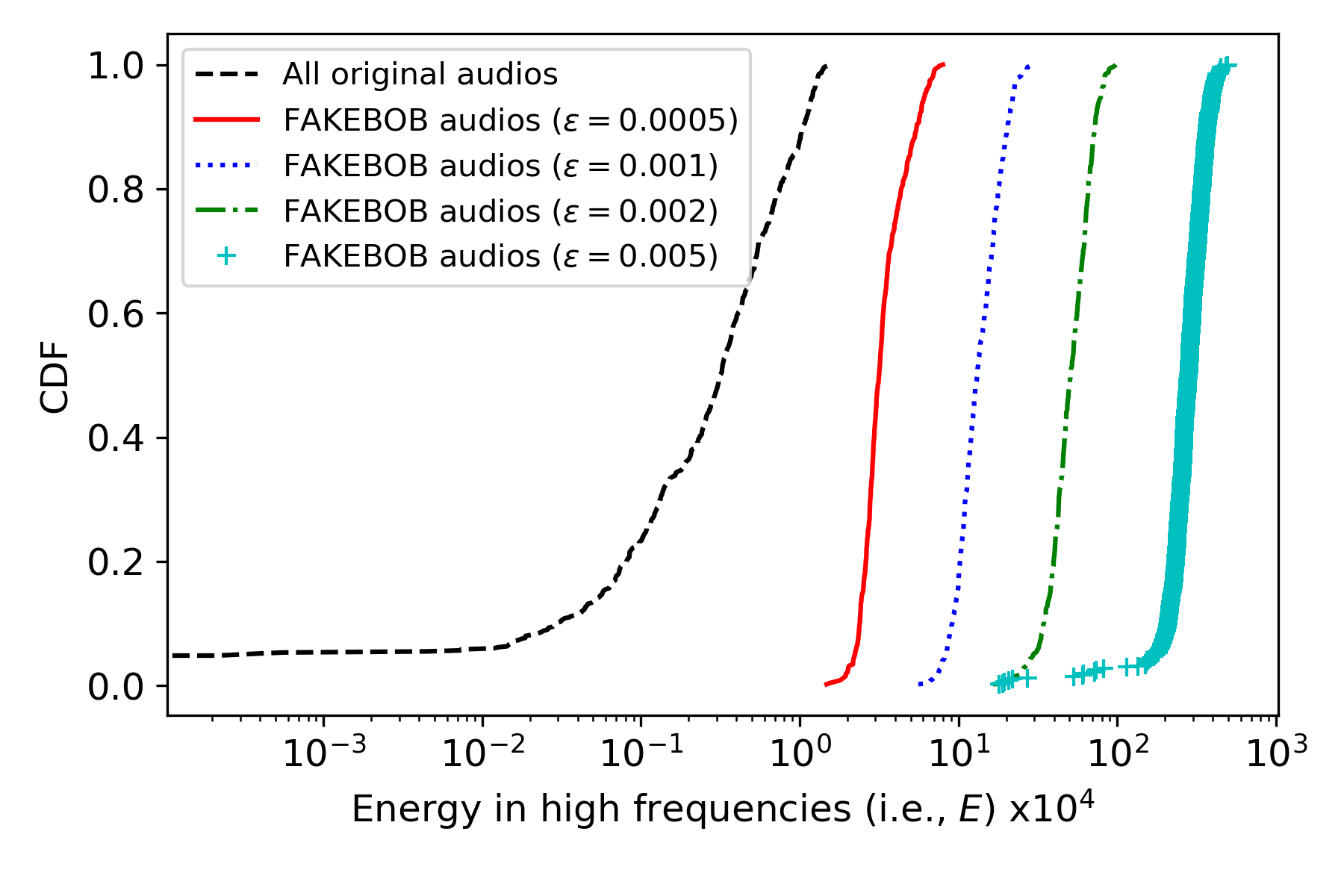}\label{fig:performance_gmm}}
      \subfigure[I-vector SV systems]{\includegraphics[width=6cm]{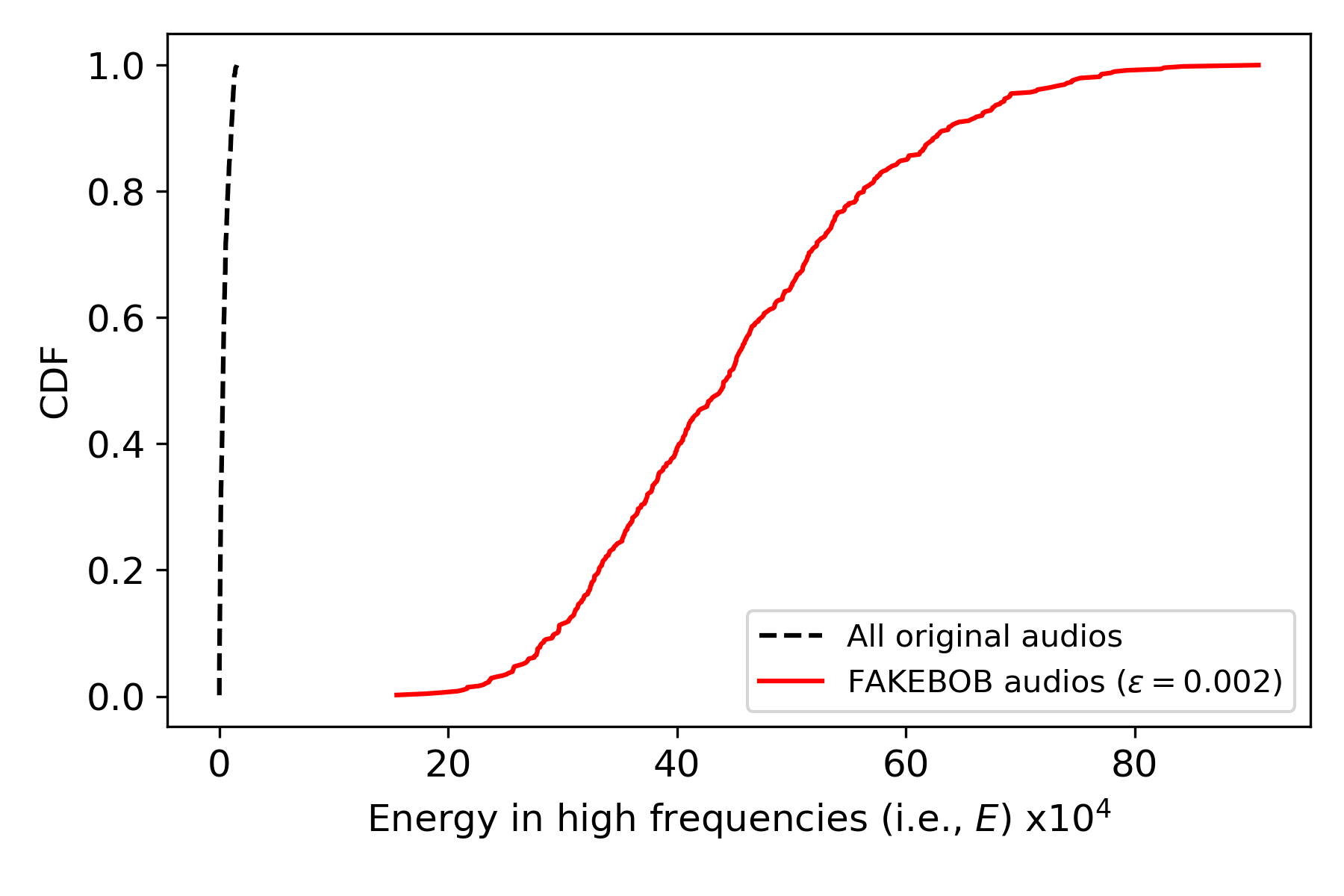}\label{fig:performance_iv}}
   }
   \caption{CDF of $E$ for original audios and FAKEBOB adversarial audios against GMM or i-vector SV systems.}
   \label{fig:performance}
   \end{center}
\end{figure*}

There are totally 600 original audios, including 500 audios from legitimate users and 100 audios from illegal users.
We randomly selected 480 ({\it i.e.}, 80\%) audios as the training data and 120 audios as the test data.
The minimum values of energy in high frequencies for the STFT of training audios ({\it i.e.}, $E$'s) were calculated,
and are with $u_E = 4.19\times 10^{-5}$ and $\sigma_E = 3.68\times 10^{-5}$.
As a result, $D = 1.52\times 10^{-4}$ based on Equation (\ref{equ:detection_threshold}).
Moreover, we study the cumulative distribution function (CDF)  $P(E \leq e)$, {\it i.e.},
the proportion of audios that are with $E$ no greater than $e$,
and plot the CDF of $E$ for training data, test data, and all original audios in Figure \ref{fig:performance_org}.
It can be seen that all three data have a similar CDF of $E$,
indicating that $E$'s in these three cases have a similar probability distribution.
Furthermore, we found that for the test data, the maximum value of $E$'s among 120 audios is $1.51\times 10^{-4}$,
which is less than $D$.
Therefore, with our experiment set, the false positive rate is zero using our proposed MEH-FEST detector,
{\it i.e.}, $P_{FP} = 0$.

Next, we consider the false negative rate and
plot the CDF of $E$ for FAKEBOB adversarial audios against GMM SV systems with different $\epsilon$,
{\it i.e.}, $\epsilon$ = 0.0005, 0.001, 0.002, and 0.005, in Figure \ref{fig:performance_gmm}.
Note that in this figure, the x-axis uses a log scale.
It can be seen that in general, when $\epsilon$ increases, the CDF of $E$ shifts to the right, indicating an overall increase of $E$.
Moreover, when $\epsilon \geq 0.001$, all values of $E$ are greater than $D$ ({\it i.e.}, $1.52\times 10^{-4}$).
When $\epsilon = 0.0005$, only one $E$ value ({\it i.e.}, $1.48\times 10^{-4}$) is less than $D$,
while all other $E$ values are greater than $D$.
Since the total number of FAKEBOB audios with four different values of $\epsilon$ is 1,872, by applying our designed MEH-FEST detector,
the false negative rate ({\it i.e.}, $P_{FN}$) is only $1/1872 \approx 0.053\%$.
We further investigated this false negative audio and found that it was a failed adversarial audio against the GMM SV system.
Therefore, if only successful adversarial audios are considered,
our detector can achieve 100\% detection rate with the experiment set.

Furthermore, we plot the CDF of $E$ for FAKEBOB adversarial audios against
i-vector SV systems with $\epsilon = 0.002$ in Figure \ref{fig:performance_iv}.
It can be clearly seen that all values of $E$ of adversarial audios are greater than $D$.
The smallest $E$ value is $1.55\times 10^{-3}$.
Therefore, the MEH-FEST detector can identify all FAKEBOB adversarial audios in these i-vector SV systems.

The experimental results indicate that our proposed MEH-FEST detector is very effective in distinguishing
between original audios and FAKEBOB adversarial audios.
Moreover, we found that it took the MEH-FEST detector averagely 3.37 milliseconds to process an input audio.
That is, the MEH-FEST method is able to provide the real-time detection.

From the theoretical perspective, if Equation (\ref{equ:sigma_D}) is applied,
we found that $\sigma_D = 2.72 \times 10^{-4}$, where $D = 1.52\times 10^{-4}$, $W = 400$, $M = 33$, and $F = 492$ from our experiments.
It can be seen that all $\sigma$'s listed in Table \ref{tab:FAKEBOB} or used in the case of i-vector SV
have a value greater than $\sigma_D$.
Therefore, our experimental results verify our theoretical analysis in Section \ref{sec:theoretical}
that when $\sigma > \sigma_D$, our designed MEH-FEST can correctly detect adversarial audios with a high probability.

\subsection{Performance of Countermeasures Against Adaptive FAKEBOB Attacks}

We evaluate the performance of defenders' countermeasures against two adaptive FAKEBOB attacks.

\subsubsection{Reducing the Perturbation Threshold $\epsilon$}
\label{sec:exp_epsilon}

As a countermeasure, an attacker would reduce perturbation threshold $\epsilon$ to avoid the detection of MEH-FEST.
However, as shown in Table \ref{tab:FAKEBOB}, when $\epsilon$ is further reduced to be less than 0.0005,
the ASR will be further reduced to be less than 61.47\%,
and meanwhile the running time will be further increased to be longer than 99 hours 5 minutes.
In this sense, our proposed detector can force FAKEBOB to reduce the attacking power.

\begin{figure}[tb]
\centerline{\psfig{figure=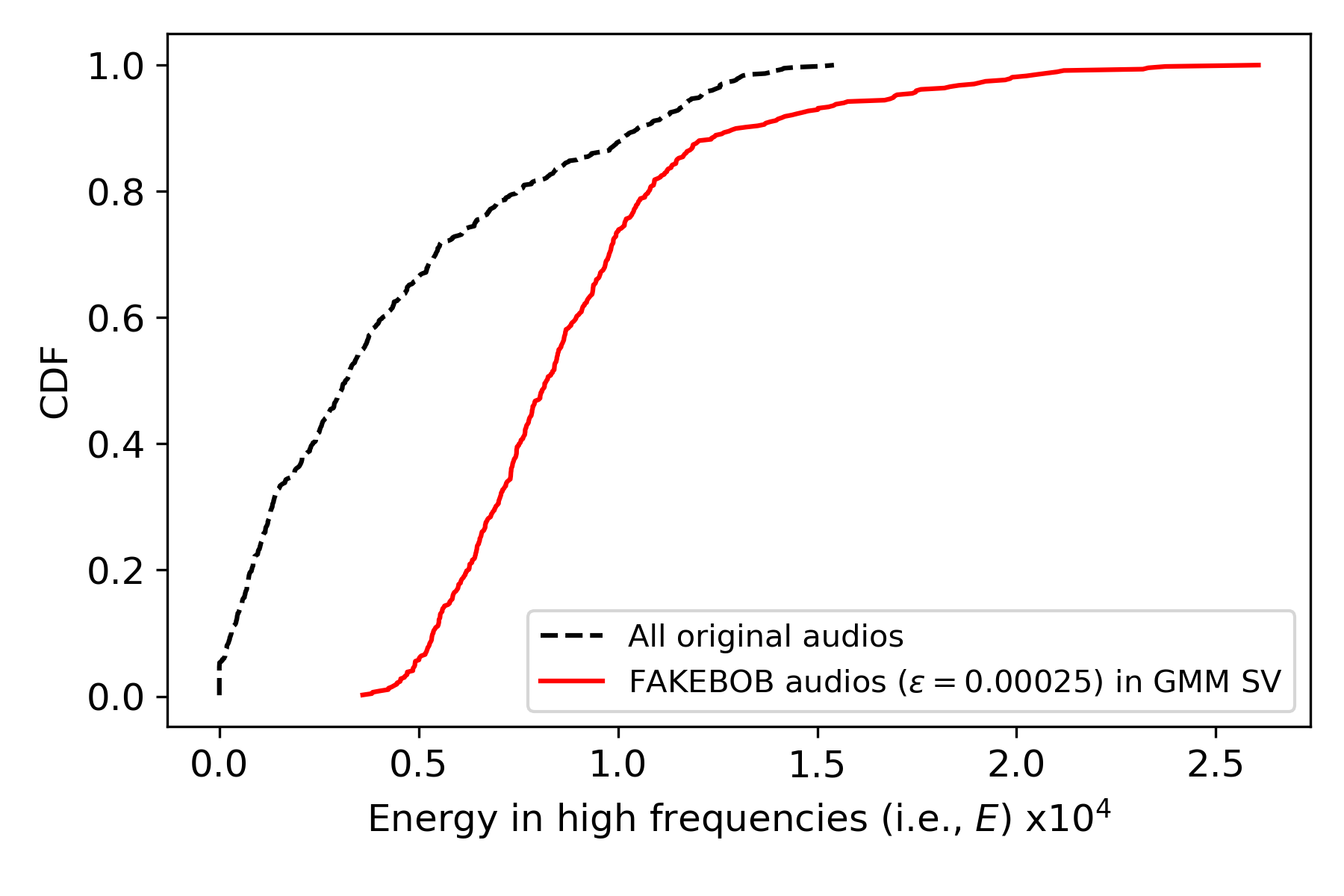, width=7.5cm}}
\caption{CDF of $E$ for FAKEBOB adversarial audios with $\epsilon=0.00025$ against GMM SV systems.}
\label{fig:epsilon_00025}
\end{figure}

We use FAKEBOB attacks with $\epsilon = 0.00025$ against GMM SV systems as an example.
When the perturbation threshold is reduced to 0.00025, the average ASR is only 42.06\%,
and the total running time increases to 143 hours 40 minutes.
On the other hand, the average $\sigma$ in $T$ in adversarial audios is reduced to $1.71 \times 10^{-4}$,
which is less than $\sigma_D$ ({\it i.e.}, $2.72\times 10^{-4}$).
We plot the CDF of $E$ for FAKEBOB adversarial audios against GMM SV systems with $\epsilon=0.00025$
in Figure \ref{fig:epsilon_00025}.
It can be seen that a majority of $E$ values are no more than $D$ ({\it i.e.}, $1.52\times 10^{-4}$).
As a result, the false negative rate of the MEH-FEST method is 93.16\%.
This verifies our theoretical analysis in Section \ref{sec:theoretical} that
when $\sigma < \sigma_D$, the adversarial audio may be able to avoid the detection of MEH-FEST.

\begin{figure}[tb]
\centerline{\psfig{figure=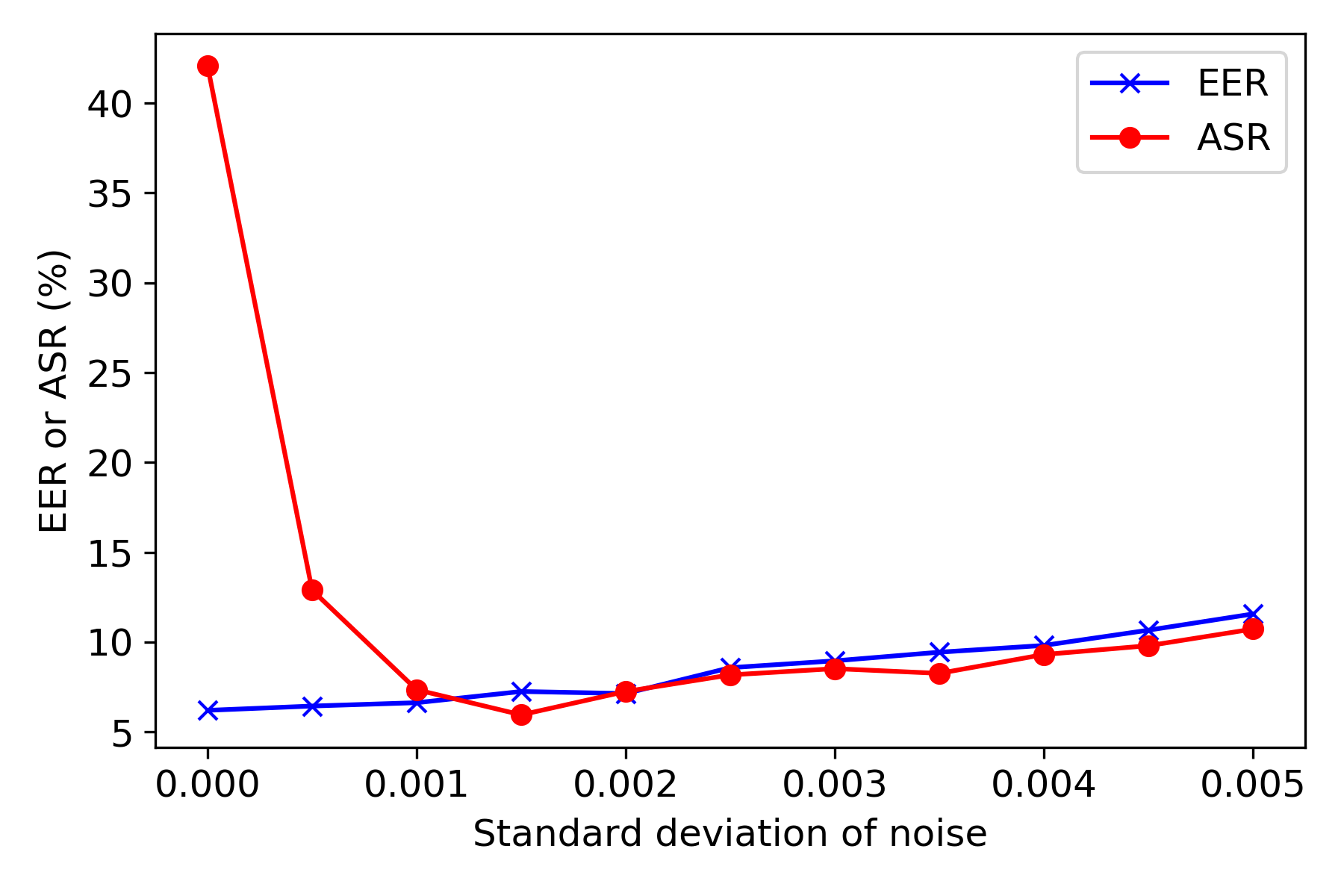, width=7.5cm}}
\caption{Noise-adding defense method against FAKEBOB attacks with $\epsilon=0.00025$ in GMM SV systems.}
\label{fig:noise_adding}
\end{figure}

\begin{figure*}[htb]
\begin{center}
    \mbox{
      \subfigure[Original audios]{\includegraphics[width=6cm]{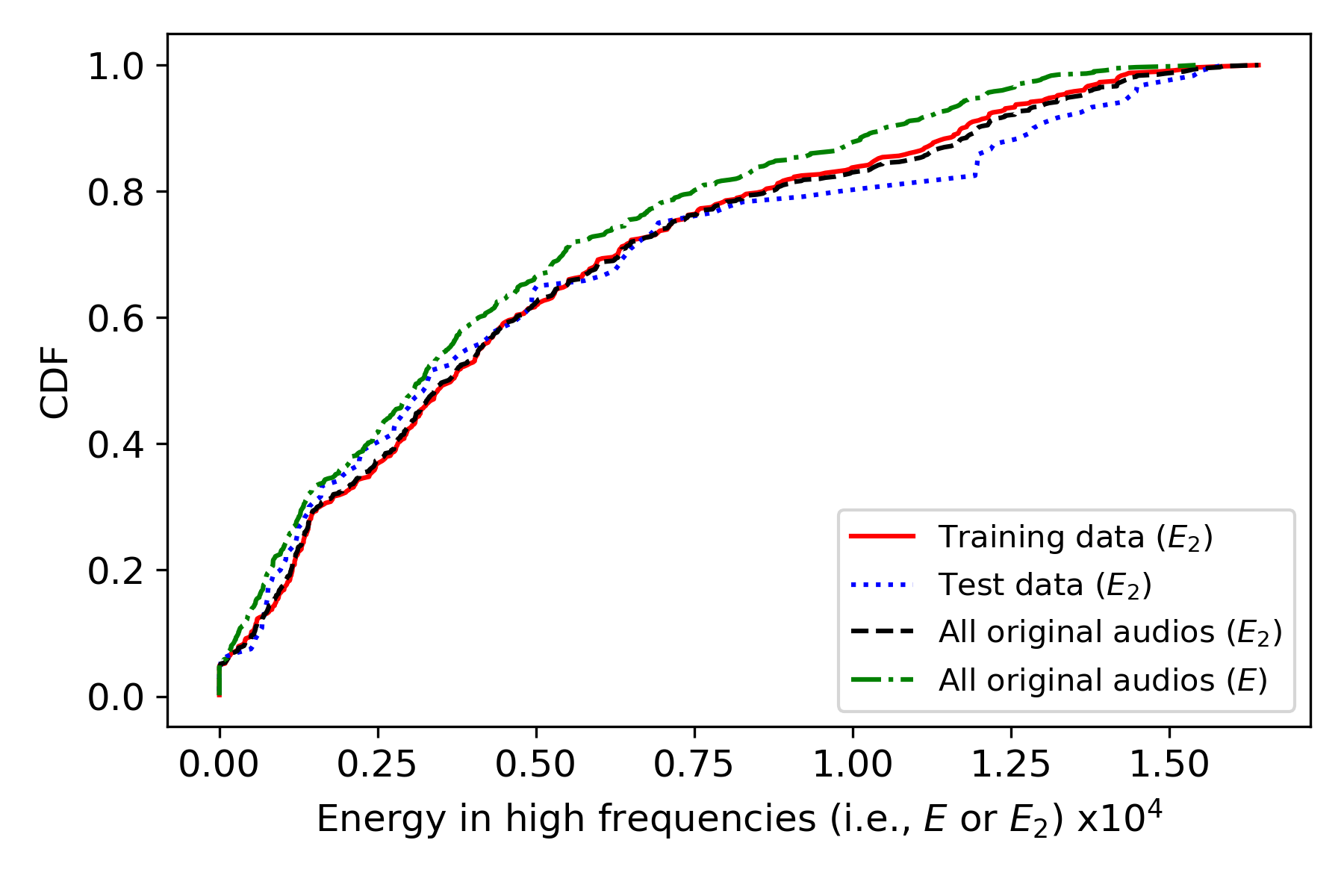}\label{fig:original_E2}}
      \subfigure[1st FAKEBOB audios]{\includegraphics[width=6cm]{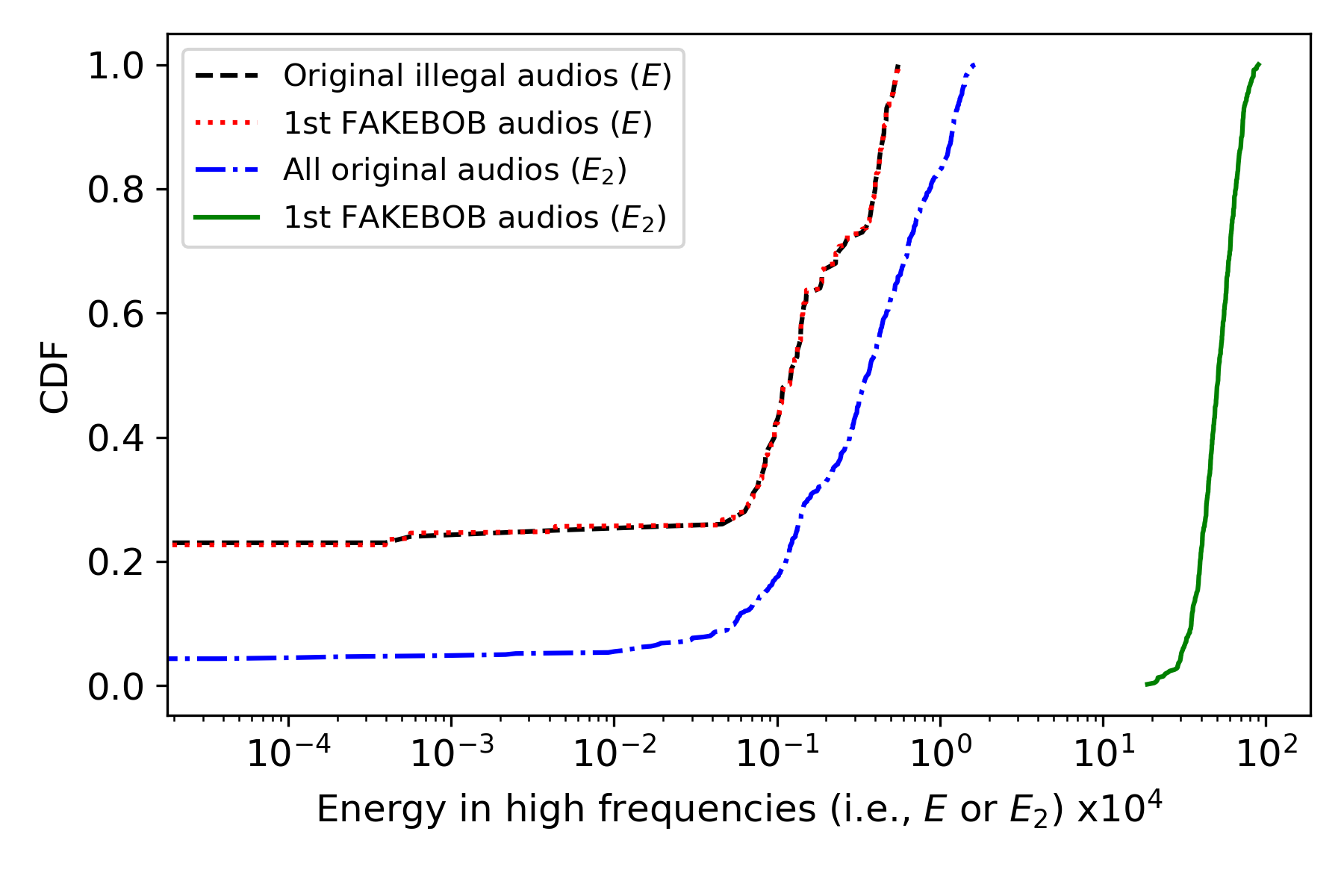}\label{fig:adv_E2}}
      \subfigure[10th and 20th FAKEBOB audios]{\includegraphics[width=6cm]{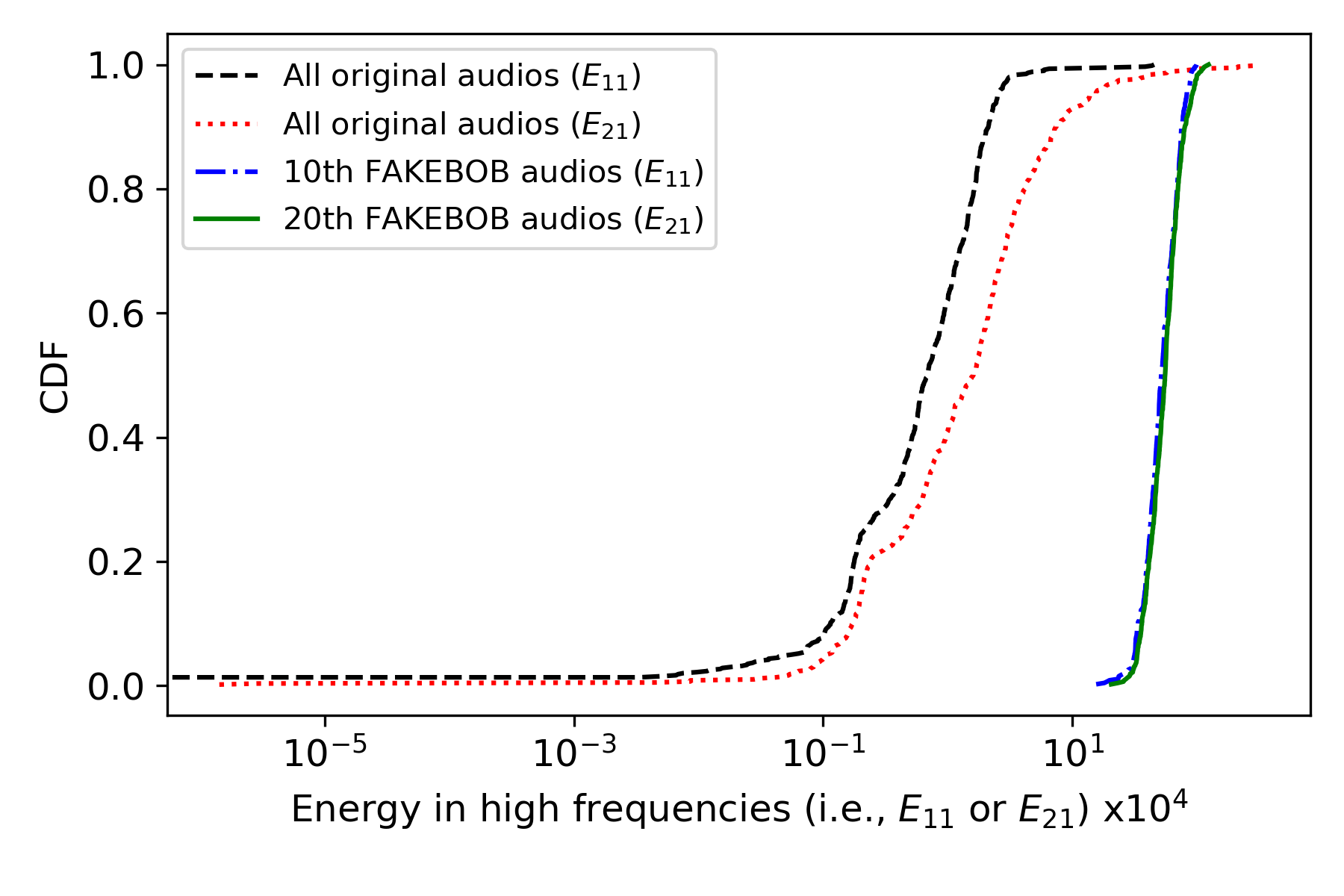}\label{fig:E11_E21}}
   }
   \caption{CDF of energy ({\it i.e.}, $E$, $E_2$, $E_{11}$, or $E_{21}$) for original audios and $n$-th FAKEBOB adversarial audios with $\epsilon = 0.002$ against GMM SV systems.}
   \label{fig:countermeasures}
   \end{center}
\end{figure*}

However, adversarial audios with a small value of $\epsilon$ are vulnerable to countermeasures
such as the noise-adding method that was proposed in our previous work \cite{Chang}.
Such a countermeasure adds small white noise to an input audio before feeding it into an SV system.
Figure \ref{fig:noise_adding} shows how adding the white noise with different standard deviation can affect
the normal operations of the GMM SV systems in terms of EER
and counteract the FAKEBOB attacks with $\epsilon=0.00025$ in terms of ASR.
Here the adversarial audios were first generated from the GMM SV systems without the noise-adding defense
and then applied to SV systems with the defense to study the ASR.
It can be seen that when the standard deviation of noise increases from 0 to 0.001,
EER increases from 6.20\% to only 6.63\%, while ASR decreases from 42.06\% to 7.35\%.
This indicates that adding the noise with a small value of standard deviation is extremely effective in defending against
FAKEBOB with a small value of the perturbation threshold $\epsilon$,
but only slightly affects the normal operations of an SV system.

\subsubsection{$n$-th FAKEBOB Attacks}

We evaluate the performance of the ($n+1$)-th MEH-FEST detection method against the $n$-th FAKEBOB adaptive attacks.
We first plot the CDFs of $E_2$ for training original audios, test original audios, and all original audios
in Figure \ref{fig:original_E2}.
It can be seen that the probability distributions for all three cases are similar.
Moreover, we also plot the CDF of $E$ for all original audios in Figure \ref{fig:original_E2}.
Comparing with $E$, the CDF of $E_2$ shifts slightly to the right, indicating that $E_2$ has a slightly larger value than $E$.
However, the difference between $E$ and $E_2$ is small.
The relative small values of $E_2$'s provide a foundation for the 2nd MEH-FEST method against the 1st FAKEBOB attacks.

In Figure \ref{fig:adv_E2}, the CDFs of $E$ and $E_2$ for both original audios and adversarial audios
from the 1st FAKEBOB attack with $\epsilon = 0.002$ against GMM SV systems are shown.
It can be seen that the CDFs of $E$ for the original illegal audios and adversarial audios are overlapping,
indicating that the 1st FAKEBOB can effectively avoid the detection of the 1st MEH-FEST detection
proposed in Algorithm \ref{algo:MEH-TEST}.
However, it also shows that the CDF of $E2$ for adversarial audios is clearly different from that for all original audios,
indicating that the 2nd MEH-FEST detection method should be able to effectively distinguish
between original audios and adversarial audios.

As an extension, we further plot the CDFs of $E_{11}$ and $E_{21}$ for original audios and adversarial audios generated by
the 10th or 20th FAKEBOB attacks with $\epsilon = 0.002$ against GMM SV systems in Figure \ref{fig:E11_E21}.
It can be seen that there is a small overlapping between two curves of the CDFs of $E_{11}$,
which means that the maximum value of $E_{11}$ from the original audios is larger than the minimum value of $E_{11}$ from
the 10th FAKEBOB attack. This would introduce a non-zero
false positive rate (FPR) or false negative rate (FNR) for our detector.
Moreover, the CDF of $E_{21}$ of original audios shifts to the right comparing with the CDF of $E_{11}$ of original audios,
whereas the CDFs of $E_{11}$ and $E_{21}$ from adversarial audios are very similar in the figure.
This indicates that when $n$ increases, there would be more overlappings between two CDFs of $E_{n+1}$
from original audio and adversarial audios generated by $n$-th FAKEBOB attacks.

Table \ref{tab:countermeasures} shows the performance of ($n+1$)-th MEH-FEST detection method
against $n$-th FAKEBOB attacks with $\epsilon=0.002$ in GMM SV systems, when $n$ varies from 0 to 50.
Note that the 0th FAKEBOB attack is the original FAKEBOB attack shown in Algorithm \ref{algo:FAKEBOB}.
It can be seen that when $n$ increases from 0 to 50, the ASR of the $n$-th FAKEBOB attack decreases from 90.24\% to 76.86\%,
whereas the total running time increases from 31 hours 3 minutes to 55 hours 5 minutes.
The value of detection threshold $D_{n+1}$ is estimated in the same way as $D$ in Equation (\ref{equ:detection_threshold}),
but $u_E$ and $\sigma_E$ are the mean and the standard deviation of $E_{n+1}$ of the training original audios.
Table \ref{tab:countermeasures} indicates that when $n \leq 10$, both FPR and FNR are zero or near zero,
showing that our proposed countermeasures can effectively defend against $n$-th FAKEBOB adaptive attacks.
However, when $n \geq 20$, while the FPR is still a small number, the FNR is very large,
which indicates that the $n$-th FAKEBOB attack can avoid the detection of the countermeasure.

\begin{table}[htb]
\begin{center}
\caption{($n+1$)-th MEH-FEST method against $n$-th FAKEBOB attacks with $\epsilon=0.002$ in GMM SV systems.}
\label{tab:countermeasures}
\begin{tabular}{|r|c|c|c|c|c|} \hline
   $n$ & ASR     & Running time    & $D_{n+1}$ & FPR    & FNR       \\ \hline
   0   & 90.24\% & 31h 3m  & 0.00015   & 0\%    & 0\%  \\
   1   & 90.66\% & 31h 54m & 0.00017   & 0\%    & 0\%  \\
   10  & 90.03\% & 33h 3m  & 0.00094   & 1.67\% & 0\%  \\
   20  & 86.81\% & 37h 48m & 0.0060    & 2.5\%  & 60.26\%  \\
   30  & 85.37\% & 39h 48m & 1.73      & 0\%    & 100\%  \\
   40  & 81.70\% & 46h 41m & 13.22     & 2.5\%  & 100\%  \\
   50  & 76.86\% & 55h 5m  & 43.65     & 5.83\% & 100\%  \\ \hline
\end{tabular}
\end{center}
\end{table}

We further manually examined the cases when $n \geq 20$ and chose a different value for the detection threshold $D^{new}_{n+1}$
as shown in Table \ref{tab:countermeasures_new_threshold}.
It can be seen from Table \ref{tab:countermeasures_new_threshold} that with the new $D^{new}_{n+1}$,
the FPR and the FNR of the ($n+1$)-th MEH-FEST method have a similar value, which results in a much better detection performance.
When $n$ is large, how to choose a proper value for the detection threshold is our future work.

\begin{table}[htb]
\begin{center}
\caption{($n+1$)-th MEH-FEST method with a different detection threshold against $n$-th FAKEBOB attacks with $\epsilon=0.002$ in GMM SV systems.}
\label{tab:countermeasures_new_threshold}
\begin{tabular}{|r|c|c|c|} \hline
   $n$ & $D^{new}_{n+1}$ & FPR     & FNR       \\ \hline
   20  & 0.0024          & 5.83\%  & 0.43\%    \\
   30  & 0.0043          & 15.83\% & 17.52\%   \\
   40  & 0.0047          & 23.33\% & 23.50\%   \\
   50  & 0.0057          & 34.17\% & 32.48\%   \\ \hline
\end{tabular}
\end{center}
\end{table}

\section{Conclusions}
\label{sec:conclusions}

In this work, we have proposed an effective detector, {\it i.e.}, MEH-FEST, against FAKEBOB adversarial attacks in SV systems.
The MEH-FEST detector was designed based on the observations that
adversarial perturbations behave like white noise and significantly affect the audio signal when the speech is absent,
especially in the high frequencies of the spectrum.
Specifically, the MEH-FEST method calculates the minimum energy in high frequencies of the STFT of an input audio signal.
We have shown through both analysis and experiments
that our designed MEH-FEST is very effective in distinguish audios before and after processed by FAKEBOB attacks.
Especially, we have demonstrated that
both false positive and false negative rates of the MEH-FEST detector are zero or approach zero in our experiments.

Moreover, we have studied the game between attackers and defenders
for the adaptive adversarial attacks and their countermeasures.
Specifically, we have designed $n$-th FAKEBOB adaptive attacks that can avoid the detection of $m$-th MEH-FEST,
when $n \geq m$.
Meanwhile, we have shown through experiments that $(n+1)$-th MEH-FEST method can be applied to
counteract $n$-th FAKEBOB attacks, especially when $n$ is not very large.

The part of source code used in this paper can be found from GitHub \cite{detection_FAKEBOB}.

As our on-going work, we plan to test our designed MEH-FEST detector on the over-the-air audios
and measure the performance in the real environment.
Moreover, we are considering to modify our proposed MEH-FEST method to
detect both replay attacks and adversarial attacks in speaker verification systems.

%

\balance


%

\end{document}